\documentclass{aa}
\usepackage{natbib}
\usepackage{lscape}
\usepackage{rotating}
\usepackage{graphicx}
\usepackage{enumerate}
\usepackage{amsmath}
\usepackage{subcaption}
\usepackage{txfonts}
\usepackage{float}
\usepackage{textcomp}
\usepackage{gensymb}
\usepackage{comment}
\usepackage{color}
\usepackage[dvipsnames]{xcolor}
\usepackage{soul}

\begin{document}

   \title{The origin of spin in binary black holes:}
   \titlerunning{The origin of spin in BBHs}
   \authorrunning{Bavera, Fragos, Qin, Zapartas, Neijssel, Mandel, Batta, Gaebel, Kimball,  Stevenson}

   \subtitle{Predicting the distributions of the main observables of Advanced LIGO}
   
    \author{Simone\,S.\,Bavera\inst{1}\fnmsep\thanks{E-mail:simone.bavera@unige.ch}, Tassos\,Fragos\inst{1}, Ying\,Qin\inst{1}, Emmanouil\,Zapartas\inst{1}, Coenraad\,J.\,Neijssel\inst{2}, Ilya\,Mandel\inst{3,2,4}, \\ Aldo\,Batta\inst{5,6,7}, Sebastian\,M.\,Gaebel\inst{2},  Chase\,Kimball\inst{8}, Simon\,Stevenson\inst{9,4}
          }

   \institute{
                Geneva Observatory, University of Geneva, Chemin des Maillettes 51, 1290 Versoix, Switzerland
                \and
                Birmingham Institute for Gravitational Wave Astronomy and School of Physics and Astronomy, University of Birmingham, Birmingham, B15 2TT, United Kingdom
                \and
                Monash Centre for Astrophysics, School of Physics and Astronomy, Monash University, Clayton, Victoria 3800, Australia
                \and
                ARC Centre of Excellence for Gravitational Wave Discovery -- OzGrav
                \and
                Instituto Nacional de Astrofísica, Óptica y Electrónica, Tonantzintla, Puebla 72840, México
                \and
                Consejo Nacional de Ciencia y Tecnología, Av. Insurgentes Sur 1582, Col. Crédito Constructor, CDMX, C.P. 03940, Mexico
                \and
                Niels Bohr Institute, University of Copenhagen, Blegdamsvej 17, 2100 Copenhagen, Denmark
                \and 
                Center for Interdisciplinary Exploration and Research in Astrophysics (CIERA), Northwestern University, 2145 Sheridan Road, Evanston, IL 60208, USA
                \and
                Centre for Astrophysics and Supercomputing, Swinburne University of Technology, Hawthorn, VIC 3122, Australia
             }

   \date{Accepted on January 21, 2020}

 
  \abstract
   {After years of scientific progress, the origin of stellar binary black holes is still a great mystery. Several formation channels for merging black holes have been proposed in the literature. As more merger detections are expected with future gravitational-wave observations, population synthesis studies can help to distinguish between them.}
   {We study the formation of coalescing binary black holes via the evolution of isolated field binaries that go through the common envelope phase in order to obtain the combined distributions of observables such as black-hole spins, masses and cosmological redshifts of mergers.}
   {To achieve this aim, we used a hybrid technique that combines the parametric binary population synthesis code COMPAS with detailed binary evolution simulations performed with the MESA code. We then convolved our binary evolution calculations with the redshift- and metallicity-dependent star-formation rate and the selection effects of gravitational-wave detectors to obtain predictions of observable properties.}
   {By assuming efficient angular momentum transport, we are able to present a model that is capable of simultaneously predicting the following three main gravitational-wave observables: the effective inspiral spin parameter $\chi_\mathrm{eff}$, the chirp mass $\mathrm{M}_\mathrm{chirp}$ and the cosmological redshift of merger $z_\mathrm{merger}$. We find an excellent agreement between our model and the ten events from the first two advanced detector observing runs. We make predictions for the third observing run O3 and for Advanced LIGO design sensitivity. We expect approximately 80\% of events with  $\chi_\mathrm{eff} < 0.1$, while the remaining 20\% of events with $\chi_\mathrm{eff} \ge 0.1$ are split into $\sim$ 10\% with $\mathrm{M}_\mathrm{chirp} < 15 \, \text{M}_\odot$ and $\sim$ 10\% with $\mathrm{M}_\mathrm{chirp} \ge 15 \, \text{M}_\odot$. Moreover, we find that $\mathrm{M}_\mathrm{chirp}$ and $\chi_\mathrm{eff}$ distributions are very weakly dependent on the detector sensitivity.}
   {The favorable comparison of the existing LIGO/Virgo observations with our model predictions gives support to the idea that the majority, if not all of the observed mergers, originate from the evolution of isolated binaries. The first-born black hole has negligible spin because it lost its envelope after it expanded to become a giant star, while the spin of the second-born black hole is determined by the tidal spin up of its naked helium star progenitor by the first-born black hole companion after the binary finished the common-envelope phase.}


   \maketitle
%

\section{Introduction}

    During the first and second observing runs O1/O2 of the advanced gravitational-wave (GW) detector network, Advanced LIGO (aLIGO) \citep[][]{2015CQGra..32g4001L} and Advanced Virgo \citep[][]{2015CQGra..32b4001A} detected 10 GWs from binary black holes (BBHs). With the third observing run O3 that just started, this number is expected to increase. In the near future, sometime around 2020, the detectors will be upgraded to reach design sensitivity and we expect the detection of hundreds of such mergers \citep[][]{2019PhRvX...9c1040A}.
    
    To date, the origin of these BBHs is still an open scientific question. Various explanations of different formation channels for merging BBHs have entered into the scientific literature \citep[see, e.g.,][for reviews]{2016ApJ...818L..22A,2016GReGr..48...95M,2018arXiv180605820M}. The most popular ones are as follows: isolated binary evolution where
    (i) the stars go through a common envelope (CE) phase due to an unstable mass transfer after the formation of the first-born black hole (BH) \citep[e.g.,][]{SmarrBlandford:1976,vdH:1976,TutukovYungelson:1993,2007PhR...442...75K,2014LRR....17....3P,2016Natur.534..512B},
    (ii) massive stars with a nonextreme mass ratio after the formation of the first-born BH goes through stable mass transfer avoiding the CE phase \citep[e.g.,][]{2017MNRAS.471.4256V,2017MNRAS.465.2092P,2017MNRAS.468.5020I},
    (iii) massive stars in close orbits experiencing strong internal mixing go through chemically homogeneous evolution and produce massive BBHs \citep[e.g.,][]{2009A&A...497..243D,2016MNRAS.458.2634M,2016A&A...588A..50M}; dynamical formation 
    (iv) in globular clusters and (v) galactic nuclear clusters where the BBHs are formed from stars not born in the same binary \citep[e.g.,][]{1993Natur.364..423S, 2000ApJ...528L..17P,2009ApJ...692..917M,2015PhRvL.115e1101R,2016ApJ...816...65A}; or 
    (vi) Lidov-Kozai resonance bringing the inner binary to merge in hierarchical triple systems \citep[e.g.,][]{SilbeeTremaine:2017}.
    All of these scenarios possess some significant uncertainties in the prediction of merger rates due to the poorly constrained underlying physics or unconstrained distributions of initial conditions. The merger rate predictions for the isolated binary evolution via the CE phase are consistent \citep{2017PhRvL.118v1101A} with the observed rate of BBH mergers of around \mbox{$\sim 24 - 112$ Gpc$^{-3}$yr$^{-1}$} \citep{2018arXiv181112940T}. The same holds for the stable mass transfer channel \citep{2019MNRAS.490.3740N}, while formation via chemically homogeneous
    evolution could yield tens of mergers per \mbox{Gpc$^{-3}$yr$^{-1}$} \citep[][]{2016MNRAS.458.2634M,2016A&A...588A..50M}. Finally, predicted rates via the dynamical formation channel are closer to the lower end of the observed range \citep[][]{2018ApJ...856...92F,2017MNRAS.469.4665P}; for example \citet{2018ApJ...866L...5R} find $4-18$ Gpc$^{-3}$yr$^{-1}$ from globular cluster in the local Universe.

    Any astrophysical BH can be fully described by only two quantities: its mass $M$ and its dimensionless spin parameter, $\mathbf{a}=c\mathbf{J}/(GM^2)$, where $\mathbf{J}$ is the angular momentum of the BH. 
    Using matched-filtering analysis, GW observations provide estimates for each of the above-mentioned quantities for both parent BHs. Although individual BH spin magnitudes and orientations are poorly constrained with present GW measurements, the effective inspiral spin parameter
    \begin{equation}
        \chi_\mathrm{eff} = \frac{M_1 \mathbf{a}_1 + M_2 \mathbf{a}_2}{M_1+M_2} \mathbf{\hat{L}} \, ,
        \label{eq:chieff}
    \end{equation}
    the mass-weighted spin of the system projected onto the orbital angular momentum $\mathbf{L}$, is reasonably well constrained \citep[][]{2019PhRvX...9c1040A}. This is explained by the fact that the leading spin-orbit-coupling term in the post-Newtonian waveforms is dominated by this parameter \citep[][]{2010PhRvD..82f4016S}. From the ten observed $\chi_\mathrm{eff}$, 8 are consistent with 0 within the 90\% credible interval while the remaining two are determined with a positive value of $\chi_\mathrm{eff}$.
    Another important quantity characterizing the waveforms is the chirp mass
    \begin{equation}
        M_\mathrm{chirp} = \frac{(M_1 M_2)^{3/5}}{(M_1+M_2)^{1/5}} \, ,
    \end{equation}
    which, to first-order approximation, determines the frequency evolution of the GW signal emitted during the BBH's inspiral phase \citep{1994PhRvD..49.2658C}. The ten observed $M_\mathrm{chirp}$ span the range of $7.9-35.7$ M$_\odot$ with a pile-up around 26 M$_\odot$.
    In addition, the luminosity distance can be measured using the GW amplitude and, assuming a cosmological model, the cosmological redshift of the merger can be inferred.
    The distributions of these parameters for a population of merging BBHs can be used to distinguish between different formation channels. As pointed out in the literature, the effective inspiral spin parameter is sensitive to the evolutionary path of BBHs \citep[see e.g.,][]{2016ApJ...832L...2R}.
    For isolated field binary channels, the spins of the two BHs are expected to be preferentially aligned with the orbital angular momentum, whereas, assuming effective exchange interaction, the spin directions of BBHs formed in dynamical environments are expected to be randomly, isotropically distributed \citep[][]{2016ApJ...818L..22A,2017Natur.548..426F}.
    
    In this study, we focus on BBHs formed through classical isolated binary evolution that go through the CE phase. The main evolutionary phases of this pathway are now summarized. At the beginning, the stars are born in a relatively wide binary where the initially more massive star, called the ``primary'', reaches the end of its main sequence first. At this stage the primary star expands its hydrogen-rich envelope past the Roche-lobe and begins transferring mass to the secondary until it loses its entire envelope, leaving a naked helium-burning star. Following wind-driven mass loss the primary collapses to form a BH. When the secondary reaches the end of its main sequence, the process repeats itself in reverse. This time, the mass transfer onto the black hole is unstable and this leads to the formation of a CE of gas around the binary \citep{1976IAUS...73...75P}. The physical details of this phase are still not fully understood \citep{2013A&ARv..21...59I}. The drag force on the BH from the envelope leads to a rapid inspiral and the dissipated orbital energy leads to the expulsion of the envelope and a decay of the orbital separation by more than two orders of magnitude. At this stage we are left with the immediate progenitor of the BBH system, namely a BH - He-star binary. Finally, the secondary eventually collapses into a BH and potential asymmetries in the core collapse may impart a kick on the newly formed BH and alter the orbit further. Eventually, due to energy and angular momentum loss from GW emission, the BBH system can coalesce into a single, more massive BH. 
    
    Previous theoretical works focused on the first few observed GW events suggest that these BBHs are consistent with having been formed through the CE formation channel \citep[][]{2017NatCo...814906S,2018MNRAS.474.2959G,2018MNRAS.481.1908K}. These authors show how, at the respective appropriate metallicity regime, the observed BH masses are produced by their binary evolution models. Furthermore, their inferred merger rates are consistent with the one obtained from GW observations. In another study in favor of the CE formation channel, \citet{2016ApJ...819..108B} carried out a detailed analysis of merger rates and found that BBHs formed though this channel should dominate the event rates in Advanced LIGO and Virgo.
    
    In the CE formation channel, the physical process determining the spin of the first-born BH is the efficiency of angular momentum (AM) transport through the evolution of the progenitor star during the red supergiant phase. From observations of astro-seismology \citep[][]{2014ApJ...796...17F,2014ApJ...788...93C}, as well as neutron star and white dwarf spins \citep[][]{2005ApJ...626..350H,2008A&A...481L..87S}, it is known that this mechanism must be efficient \citep[][]{1999A&A...349..189S,2002A&A...381..923S,2019MNRAS.485.3661F}. Thus, upon expansion, the initial angular momentum is mostly transported to the outer layers of the star which are subsequently lost due to Roche-lobe overflow mass transfer and wind mass loss. This leads to very slowly spinning BHs ($a_1 \simeq 0$) as was shown by \citet{2018A&A...616A..28Q} \citep[see also][]{2019ApJ...881L...1F}.
    Assuming efficient AM transport, the angular momentum of the second-born BH is mainly determined by the net effect of the stellar winds and the tidal interaction of the BH-He-star binary system. This is because any initial or acquired rotation during the evolution of the secondary is erased through mass transfer and wind mass loss by the time it becomes a He-star. Several studies attempted to model the last evolutionary phase of this channel and derived constraints on the spin using analytical arguments and semi-analytical calculations \citep[][]{2016MNRAS.462..844K,2017ApJ...842..111H,2018MNRAS.473.4174Z}. These studies found out that around half of the secondary BHs have zero spin and the other half are maximally spinning. When using detailed models to simulate the binary evolution and the stellar structure of the two components, \citet{2018A&A...616A..28Q} did not reproduce this prediction of a bi-modal distribution of spins. Both \citet{2017ApJ...842..111H} and \citet{2018MNRAS.473.4174Z} results are based on the approach outlined in \citet{2016MNRAS.462..844K}. Compared to the detailed binary simulations of \citet{2018A&A...616A..28Q}, these authors did not model self-consistently the orbit evolution of the binary due to the combined effects of tides and stellar winds, which in most cases leads to the widening of the orbit.  Even when tides are initially efficient at synchronizing the spin of the helium star to the orbit, such wind-driven orbital widening can lead to tidal decoupling. Ignoring this effects underestimates the impact of stellar winds on the final spin of the second-born BH. Moreover they used approximate timescales for the process of tidal synchronization that do not take into account changes in the structure of the star during its lifetime and assumed that tides allow the He-star to remain tidally locked indefinitely. These approximations lead to results that disagree with what is found in our detailed binary simulations. \citet{2017arXiv170607053B} used parametric binary population synthesis models, which share all the same approximations as the studies discussed above, to compare three different prescriptions for the efficiency of AM transport. They found that efficient AM transport is favored, as it results to distributions of $\chi_\mathrm{eff}$ and BH masses qualitatively consistent with observations, while inefficient AM transport would lead to rapidly  spinning BHs which are currently not observed by aLIGO.
 
    In this paper, we present a model capable of predicting simultaneously the spin, mass and redshift distributions of coalescing BBHs formed from isolated field binaries that go through the CE phase. This aim is achieved by combining the parametric binary population syntheses code COMPAS with the detailed MESA stellar structure and binary evolution simulations.
    The study is structured as follows. In Sec. \ref{sec:method} we explain the methods used to generate a Monte-Carlo population of isolated field BBHs and how we take into account the redshift and metallicity dependence of the star formation rate and the observational selection effects. In Sec. \ref{sec:results} we present our main results and make detailed predictions for observing runs of the future GW detector network, distinguishing three regions of the BBH parameter space. Our model is able to successfully reproduce the observed $\chi_\mathrm{eff}$ distribution providing strong support for the CE channel being the dominant formation channel. We then discuss our results in Sec. \ref{sec:discussion} where we compare our work to the current literature and demonstrate the importance of detailed binary evolution calculations. Finally, the conclusions of our work are given in Sec. \ref{sec:conclusion}.

    \begin{figure*}
        \centering
        \includegraphics[width=18cm]{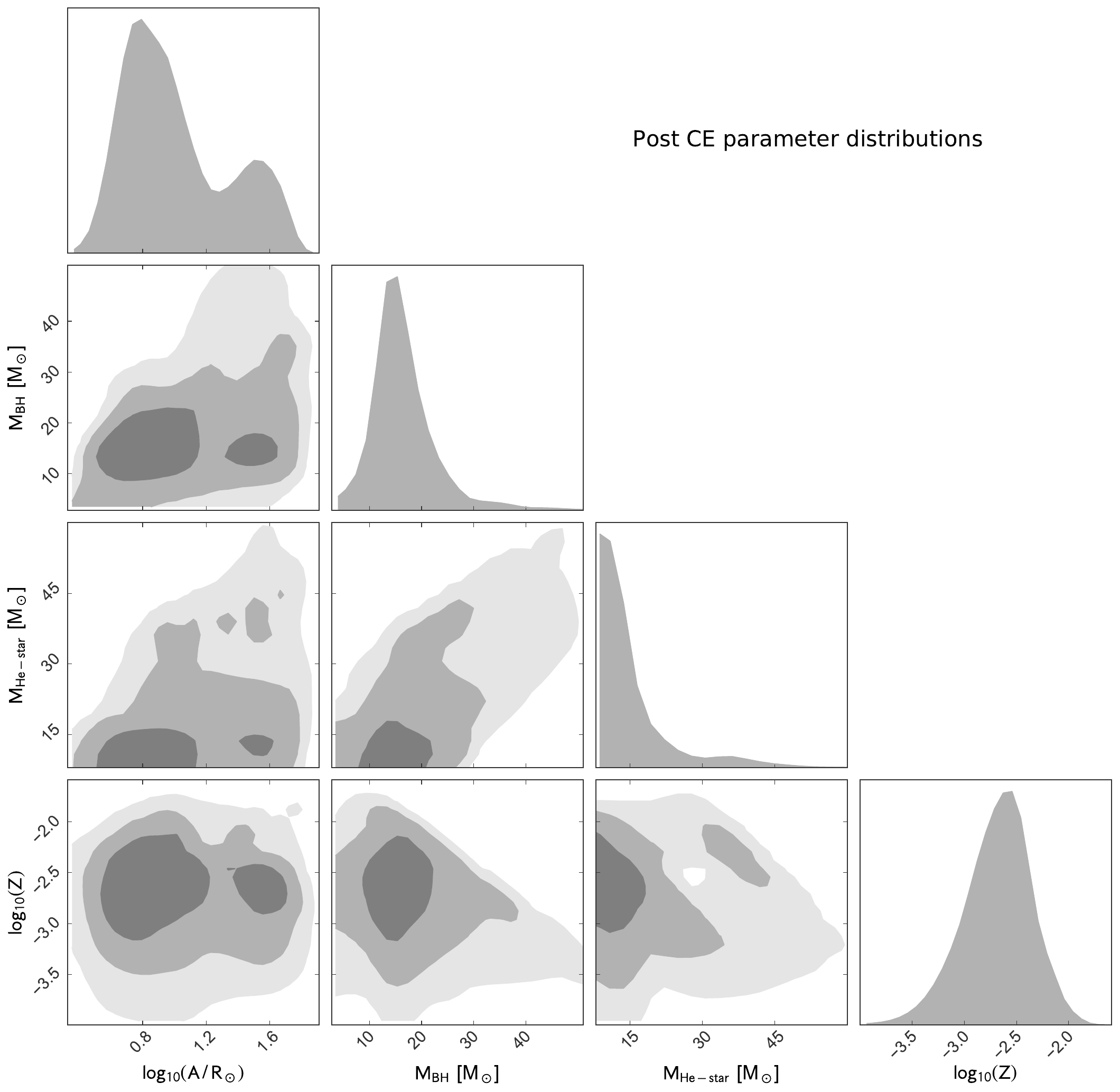}
        \caption{Parameter distributions of the binary population after the CE phase. These BH -- He-star binaries include systems that are going to form BH-NS binaries and BBHs with GW inspiral timescales bigger than the Hubble time. We show the distributions of orbital separation $\mathrm{A}$, first-born BH mass $\mathrm{M}_\mathrm{BH}$, He-star mass $\mathrm{M}_\mathrm{He-star}$ and metallicity $\mathrm{Z}$ weighted by the integrated redshift- and metallicity-dependent SFR over the cosmic time (see Eq. \ref{eq:ISFR}). The lighter shades represent larger contour levels, 68, 95 and 99\%, respectively, constructed with pygtc \citep{Bocquet2016}.}
        \label{fig:postCE}
    \end{figure*}

\section{Method}\label{sec:method}

    We use the parametric binary population synthesis code COMPAS \citep[e.g.,][]{2017NatCo...814906S,2019ApJ...882..121S,2017IAUS..325...46B,2018MNRAS.477.4685B, 2018MNRAS.481.4009V,2019MNRAS.490.3740N} to evolve isolated stellar binaries until the formation of BH - He-star systems, namely the immediate progenitors of BBHs. In Sec. \ref{sec:COMPAS} we briefly describe the COMPAS-model assumptions used in the simulation. Since we are not interested in a parameter study, we specifically picked a model capable of reproducing a BBHs merger rate which is in agreement with the observed one from  \citet{2019PhRvX...9c1040A}. For the last step of the evolution, which we consider to be the one that determines the spin distribution of the secondary BH, we use the stellar structure and evolution code MESA \citep[][\citeyear{2013ApJS..208....4P}, \citeyear{2015ApJS..220...15P}, \citeyear{2018ApJS..234...34P}, \citeyear{2019ApJS..243...10P}]{2011ApJS..192....3P} to simulate the evolution of the binary systems. Assuming that the first-born black hole can be treated as a point-like particle, this approach allows us to track the angular momentum profile evolution of the He-star until the formation of the secondary BH (see Sec. \ref{sec:MESA}). In Sec. \ref{sec:collapse}, we explain in detail how we treat the collapse of the He-stars into BHs. Finally, taking into account the redshift dependence of metallicity, star-formation rate (SFR) and aLIGO sensitivity, we can distribute the population across cosmic time and compute the expected detection rate (see Sec. \ref{sec:R_det}).

\textbf{\subsection{COMPAS model assumptions}\label{sec:COMPAS}}
    We use the results from the simulations of \citet{2019MNRAS.490.3740N} and here we only highlight the main physical assumptions.  We assume that the underlying stellar population spans the mass range \mbox{$0.01 \, \text{M}_\odot < m_1 < 200 \, \text{M}_\odot$} following the initial mass function of \citet{10.1046/j.1365-8711.2001.04022.x}. The mass distribution of the less massive secondary star is given by \mbox{$m_2 = m_1 q_0$} where $q_0$ is the initial mass ratio ($0<q_0<1$) drawn from a flat distribution \citep[][]{2012Sci...337..444S}. We are interested in binaries with a primary star that ends up forming a BH, thus we restrict the initial mass distribution of primary masses between $5 \, \text{M}_\odot < m_1 < 150 \, \text{M}_\odot$. This means we only model a fraction $f_\mathrm{corr}$ of the underlying stellar population mass. We calculate this by assuming a binary fraction of 70\% \citep{2012Sci...337..444S}, see Appendix \ref{app:Normalization}. We assume that, at formation, binaries are distributed uniformly in log-orbital separation restricted to $0.1 < A / \text{AU} < 1000$ \citep{1983ARA&A..21..343A} and have zero eccentricity. We assume that all these distributions are independent from each other as well as independent of metallicity. For the metallicity distribution of binaries we divide uniformly in 30 bins the log-metallicity range $Z \in [0.0001,0.0349]$. We then evolve three million binaries per metallicity bin $\Delta Z_j$ with a total of star forming mass on the order of $M_{\mathrm{sim},\Delta Z_j} = 6.5 \cdot 10^{7} \, \mathrm{M}_\odot$. 
    
    COMPAS evolves stars according to the stellar models of \citet{1998MNRAS.298..525P} and uses analytical fits of these models to rapidly evolve binaries \citep[][]{2000MNRAS.315..543H,2002MNRAS.329..897H}. We adopt wind mass loss rates as prescribed by \citet{2000MNRAS.315..543H} for stars with effective temperatures smaller than $12,500 \, \mathrm{K}$, and for hotter stars the winds of \citet{2001A&A...369..574V} as implemented by \citet{2008ApJS..174..223B}. If stars during their evolution cross the Humphrey-Davidson limit \citep{1994PASP..106.1025H} and enter a region of the Hertzsprung-Russel diagram in which no stars are observed, we apply an additional wind mass loss rate of $1.5 \cdot 10^
    {-4} \, \mathrm{M}_\odot \, \mathrm{yr}^{-1}$ \citep[][]{2010ApJ...714.1217B}.
    
    When the primary star reaches the end of its main sequence, the star expands and loses its entire envelope through Roche-lobe overflow. In binaries where the first mass transfer episode is stable, the companion can accrete some mass with an efficiency that we assume depends on the ratio of the thermal timescales of the two stars \citep[][]{2002MNRAS.329..897H,2015ApJ...805...20S,2019MNRAS.490.3740N}, while the mass not accreted by the other star leaves the system carrying away the specific angular momentum of the accretor. Eventually, the envelope of the primary is stripped, leaving a naked helium burning star which, following wind-driven mass loss, collapses into a BH. The star collapses into a point-like BH following the ``delayed'' model of \citet{2012ApJ...749...91F}. 
    
    When the secondary reaches the end of its main sequence the process repeats in reverse and the mass transfer between the BH and the He-star can be either dynamically stable or unstable. Since we focus only on the subchannel that goes through the CE phase, we consider exclusively systems with dynamically unstable mass transfer which produces a non co-rotating CE of gas engulfing the binary. 
    This represents only a subset of all merging binary black holes in the models of \citet{2019MNRAS.490.3740N}. While uncertainties in the stability of mass transfer could reduce the importance of the non-CE channel, a self-consistent variation of the assumptions regarding mass transfer stability would also change the population of systems that evolve through a CE phase.  This analysis is beyond the scope of this work, but could impact, in particular, our overall rate predictions for BBH mergers, which should be compared directly to \citet{2019MNRAS.490.3740N}.

    COMPAS uses the classical energy $\alpha_{CE}-\lambda$ formalism \citep[][]{1984ApJ...277..355W,1990ApJ...358..189D,2000A&A...360.1043D,2010ApJ...716..114X} to parameterize the uncertainties in the physics of the CE phase. During this phase the two stars spiral in due to friction with the envelope. The loss of orbital energy can heat up and expand the envelope. The efficiency of this energy transfer is parameterized by the $\alpha_{CE}$ parameter which can vary \citep[][]{1988ApJ...329..764L}. We assume that all of the dissipated orbital energy goes into expelling the envelope, $\alpha_{CE}=1$. The $\lambda$ parameter, which characterizes the binding energy of the CE, depends on the structure of the donor's envelope \citep[][]{1990ApJ...358..189D}. We chose our $\lambda$ according to the fits of \citet{2010ApJ...716..114X} as implemented by \citet{2012ApJ...759...52D}. 

    Within the CE subchannel, we distinguish two different scenarios for donor stars which are on the Hertzsprung-gap (HG). In the optimistic scenario we apply the usual $\alpha-\lambda$ prescription to evolve these systems. In the pessimistic scenario we assume that Hertzsprung-gap stars have not yet developed a sufficiently sharp density gradient at the core-envelope boundary to allow the inspiral during the CE to stop. Thus any CE event from donors in this evolutionary phase results in a merger which reduces the BBH merger rate \citep{2007ApJ...662..504B}. In this paper we present the results for the latter scenario. Both scenarios yield a similar distribution of spins, but the pessimistic one predicts fewer low-mass BHs compared to the optimistic. This is because a greater fraction of the total post-main-sequence expansion occurs during the HG for high-metallicity stars \citep{2010ApJ...725.1984L}. Therefore, forbidding the channel with HG CE donors has a particularly strong effect at high metallicity, which yields lower-mass BHs due to metallicity-enhanced stellar winds.
    
    In Fig.~\ref{fig:postCE} we show the distributions of orbital separation, first-born BH mass, He-star mass and metallicity of our BH-He-star binaries after the CE phase. These distributions are weighted by the redshift- and metallicity-dependent star-formation rate (SFR) integrated over the cosmic time, see Eq. (\ref{eq:ISFR}) in Appendix \ref{app:DetectionRate}. We note that these distributions include all binaries, including those systems that are going to become BH-NS binaries and BBHs with GW inspiral timescales longer than the Hubble time. After the CE phase the orbital separations are no longer uniformly log-distributed and most of the first-born BHs have masses smaller than 30 M$_\odot$. Moreover, we see that formation metallicities of progenitors of merging compact-object binaries follow a skewed log-normal distribution. This is because the mean metallicity of the Universe decreases as a function of the look-back time and the star-formation rate peaks at a redshift $\sim 2$ , that is a look-back time of $\sim 10.5$ Gyr, where most of the binaries are formed. These distributions are used as an initial condition for our detailed modeling.

\textbf{\subsection{MESA model assumptions}\label{sec:MESA}}
    We perform detailed stellar structure and binary evolution calculations that take into account wind mass loss, internal differential rotation of the He-star and tidal interaction between the BH and the He-star. These simulations\footnote{The detailed list of parameters used for the simulations can be found at http://mesastar.org/results.} are based on the work of \citet{2018A&A...616A..28Q} and are adapted for MESA r-10398.
    
    Stellar winds play an important role in binary evolution. Here, we take a slightly different approach compared to \citet{2018A&A...616A..28Q}, and we follow the wind prescriptions outlined in \citet{2010ApJ...714.1217B}, which is the same as the ones used in COMPAS. Namely, for helium stars we adopt a wind mass-loss rate of 
        \begin{equation}\label{winds}
        \frac{dM}{dt} = 10^{-13}\left(\frac{L}{L_\odot}\right)^{1.5}\left(\frac{Z}{Z_\odot}\right)^{0.86} \, \, \, \mathrm{M}_\odot \, \mathrm{yr}^{-1} \, ,
    \end{equation}
    where $L$ and $Z$ are the star's luminosity and metallicity, respectively. This prescription is a combination of \citet{1998A&A...335.1003H} and \mbox{\citet{2005A&A...442..587V}} and takes into account He-star winds clumping and a strong dependence on the metallicity. Furthermore, we adopt $Z_{\odot}$ = 0.017 as solar metallicity \citep[][]{1998SSRv...85..161G}.

    Tidal forces are responsible for synchronising the spin of the He-star with the orbit. We assume that the CE ejection leaves a circular binary, and the system remains circular during He-star evolution. It has been suggested that dynamical tides are dominant for stars with a radiative envelope and a convective core \citep[][]{1977A&A....57..383Z,1981A&A....99..126H}. The strength of the interaction depends on the ratio of the stellar radius $R$ to the orbital separation $A$. The timescale for synchronization is defined as
    \begin{equation}
        \frac{1}{T_\mathrm{sync}} = 3 E_2 \left(1+q\right)^{5/6} \frac{q^2}{r_\mathrm{g}^2} \left( \frac{GM}{R^3} \right)^{1/2} \left( \frac{R}{A} \right)^{17/2} \, ,
        \label{eq:tides}
    \end{equation}
    where the He-star has mass $M$, radius $R$ and moment of inertia $I$, $r_g$ given by $r_\mathrm{g}^2=I/(MR^2)$ is the dimensionless gyration radius of the He-star, $q$ is the mass ratio of the BH mass to the He-star mass, and $E_2$ is the second order tidal coefficient. We take the new fitting formula of $E_2$ as suggested by \citet{2018A&A...616A..28Q} for He-stars
    \begin{equation}\label{E2}
        E_2 = 10^{-0.93} \left( \frac{R_{\rm conv}}{R} \right)^{6.7} \, ,
    \end{equation}
    where $R_{\rm conv}$ is the radius of the convective core \citep[see Appendix A in][for an in-depth discussion of $E_2$]{2018A&A...616A..28Q}. We highlight here that a variation of the implementation of tides is used \citep{2019ApJ...870L..18Q}. Instead of the standard tides prescription in MESA \citep{2015ApJS..220...15P} that synchronize the whole star, the tides here only operate on the radiative regions. However it has been verified that this slight variation has a very small impact on our results.
    
    Rotational mixing and angular momentum transport are treated as diffusion processes \citep[][]{2000ApJ...528..368H,2005ApJ...626..350H}, which mainly involve the effects of Eddington-Sweet circulation, the Goldreich-Schubert-Fricke instability, and secular as well as dynamical shear mixing. In addition, diffusion element mixing is included with an efficiency parameter of $f_c$ = 1/30 \citep[][]{1992A&A...253..173C,2000ApJ...528..368H} for all processes above. Furthermore, an efficient angular momentum transport mechanism (i.e., Tayler-Spruit dynamo: \citet{1999A&A...349..189S,2002A&A...381..923S}) is included. For comparison, we also ran a small grid without the Tayler-Spruit dynamo and found that there is a negligible impact on our results. More details on this can be found in the discussion. Furthermore, efficient AM transport allows us to assume that all He-stars emerging from the CE phase are initially not rotating. This is because any initial or acquired rotation during the evolution of the secondary is erased by mass transfer and wind mass loss by the time it becomes a He-star.
    
    Running these simulations for all binary systems computed by COMPAS is computationally too expensive. Therefore we run a grid that allows us to infer through interpolation the six parameters we are interested in, namely: the He-star mass before the supernova, the carbon-oxygen (CO) core mass pre-supernova, the resultant second-born BH mass, the orbital period pre-supernova, the lifetime of the BH-He-star binary system (from the expulsion of the CE to the collapse of the He-star) and the spin of the second-born BH, as a function of the initial parameters of the BH - He-star binary: initial mass of the first-born BH, $m_\mathrm{BH}$, initial mass of the He-star, $m_\mathrm{He-star}$, initial orbital period, $p$, and He-star metallicity, $Z$. In order to optimally construct our grid, we first randomly generate 3\,000 points to cover the parameter space spanned by the binaries after the CE phase, namely, $m_\mathrm{BH}\in [2.5\, \text{M}_\odot,60\, \text{M}_\odot]$, $m_\mathrm{He-star}\in[2.5\, \text{M}_\odot,89\, \text{M}_\odot]$, $p\in[0.05 \, \text{days},8.5 \, \text{days}]$ and $Z\in[0.0001, 0.0349]$, to which we add 1,500 points drawn from a kernel density estimator (KDE) trained with the post CE phase parameters of the synthetic population. In Appendix \ref{app:Interpolation} we explain how we try to minimize linear interpolation errors by running more simulations where the interpolator is under-performing. The accuracy of the linear interpolator at each step is verified conducting 50 ``leave 5\% of the sample out'' tests. We run a total of 18,000 simulations and show how the median relative errors stabilize at 0.01\% and 0.04\% for the He-star mass pre-supernova and resultant BH mass respectively, 0.20\% for the CO core mass of the He-star, 0.01\% for the orbital period, 0.04\% for the lifetime of the BH-He-star binary and 0.41\% for the log-spin of the second-born BH. In the Appendix we also show the spread of the relative errors. If in the 50 leave 5\% of the sample out tests we also count non-fittable points, such as remote points at the boundary of the parameter space, we find the following percentages of test systems that have relative errors above 10\% in the estimated quantities: 5\% and 8\% of the He-star mass pre-supernova and resultant BH mass, 8\% of the CO core mass of the He-star, 6\% of the orbital period, 6\% of the lifetime of the BH-He-star binary and 17\% of the log-spin of the second-born BH .

\textbf{\subsection{Core-collapse physics}\label{sec:collapse}} 
    Black holes are formed during the core collapse of massive stars and, in some cases, their formation may be accompanied by supernova explosions. The collapse occurs when the stellar core begins to contract under its own weight without being able to trigger any more nuclear burning in its iron core. This leads eventually to electron capture and dissociation of the core elements into alpha particles. The first process removes the degeneracy pressure support of the core while the second removes the thermal support. These two mechanisms combined accelerate the collapse until the core reaches nuclear densities and neutron degeneracy pressure halts the collapse. This sudden halt produces a bounce shock moving out of the core. The shock-wave moves outwards until it deposits all its energy into the surrounding layers. A supernova explosion occurs if the deposited energy can overcome the ram pressure of the infalling stellar material. A fraction, $f_{fb}$, of the material ejected by the supernova then falls back onto the stellar remnant. If the remnant is massive enough, neutron degeneracy pressure fails to halt the collapse and a black hole is formed. Moreover, the most massive stars can directly overcome the neutron degeneracy pressure when the collapse starts and implode to form a black hole. For a thorough review of our current understanding of the core-collapse process see for example \citet{2007PhR...442...38J}.
    
    We use \citet{2012ApJ...749...91F} delayed supernova prescription to model how much baryonic remnant mass is left behind after the collapse of the secondary star. This differs from their rapid prescription which produces a mass gap between BHs and neutron stars by assuming a  strong convection which allows instabilities to grow quickly after the core bounce, producing more energetic SN explosion. The two prescriptions are not expected to lead to significant differences in the detected BBH merger rate \citep[][]{2016ApJ...819..108B} as the population is dominated, due to aLIGO's selection effects, by more massive BHs.
    
    Using the delayed prescription, we calculated the fraction of the star that collapses to form the BH, and we assume that any remaining outer layers that do not collapse are instantaneously ejected. In order to estimate the spin of the resulting BH we follow the framework described in \citet{2019arXiv190404835B}. We assume that there is no pressure stopping or slowing down the collapse. We can think of the star mass distribution $M(r)$ as a collection of shells with mass $m_{\rm shell}$ and angular frequency $\Omega_{\rm shell}$, that falls one by one onto the center of the star. We assume that at the center, the shells up to 2.5 M$_\odot$ collapse directly to form a black hole conserving their angular momentum and mass. Once a shell reaches the BH's event horizon, it is accreted by it. The amount of angular momentum of the infalling material determines the properties of the accretion flow. Low angular momentum material collapses directly onto the BH transferring its entire mass and angular momentum to the BH, while material with enough angular momentum can create a disk around it. The maximum amount of angular momentum the disk material can give to the BH is determined by the specific angular momentum at the innermost circular orbit (ISCO) around the BH \citep[][]{1972ApJ...178..347B},
    \begin{equation}
        j_{\rm isco} = \frac{GM_{\rm BH}}{c} \frac{2}{3^{3/2}}
        \left(1+2 \left(\frac{3c^2 r_{\rm isco}}{G M_{\rm BH}}-2 \right)^{1/2} \right) \, ,
    \end{equation}  
    where $r_{\rm isco}$ is the radius at ISCO for prograde equatorial orbits,
    \begin{equation}        
        r_{\rm isco} = \frac{GM_{\rm BH}}{c^2}\left(3+z_2 - \left( \left(3 - z_1 \right)\left(3 + z_1+ 2z_2 \right)\right)^{1/2}\right)  \, ,
    \end{equation}
     with $z_1=1+(1-a^2)^{1/3}((1+a)^{1/3}+(1-a)^{1/3})$ and $z_2 = (3a^2+z_1^2)^{1/2}$ where $a$ is the spin of the BH. Assuming that the disk formed from the collapse of a shell is accreted before the next shell collapses, as the viscous timescale of the disk is shorter than the dynamical timescale of the collapsing shells, we can evolve the BH's mass and spin as it accretes material through the accretion disk. Each mass shell then contributes to the angular momentum of the BH by
    \begin{equation}
        J_{\rm shell} = \int_0^{\theta_{\rm disk}} m_\mathrm{shell} \Omega_\mathrm{shell}(r) r^2 \sin^3 \theta \, d\theta
        + \int_{\theta_{\rm disk}}^{\pi/2} m_\mathrm{shell} j_{\rm isco} \sin \theta \, d\theta \, ,
        \label{eq:Jshell}
    \end{equation}
    where the disk formation angle is given by 
    \begin{equation}
        \theta_{\rm disk} = \arcsin \left( \left(\frac{j_{\rm isco}}{\Omega_\mathrm{shell}(r)r^2} \right)^{1/2} \right) \, 
    \end{equation}
    and if the argument of $\arcsin$ exceeds 1, there is no disk formation. The first term in Eq. (\ref{eq:Jshell}) represents material with low angular momentum that collapses directly onto the BH, while the second term corresponds to the material that forms the accretion disk with mass $m_{\rm disk}=m_{\rm shell} \cos \theta_\mathrm{disk}$. The mass-energy accreted from the disk onto the BH is $\Delta M_{\rm disk} = m_{\rm disk} (1-2 GM_{\rm BH}/(3 c^2 r_{\rm isco}))^{1/2}$ and the accreted angular momentum is $\Delta J_{\rm disk} = m_{\rm disk} j_{\rm isco} $ \citep[][]{1970Natur.226...64B,1974ApJ...191..507T}. When treating the accretion of the portions of the shell that collapse directly onto the BH, we take into account 10\% of baryonic mass loss through neutrinos \citep{2012ApJ...749...91F}. 
    
    In our population synthesis study we neglect the effects of pair-instability supernovae (PISNe) and pulsational pair-instability supernovae (PPISNe). Both events are caused by the production of electron-positron pairs in the cores of very massive stars. In a PISN, pair production leads to a drop in the radiation pressure support in the core, causing the core to contract and the core temperature to increase. This results in explosive oxygen burning which reverses the collapse, unbinding the star. A PPISN is similar but the release of energy is insufficient to completely disrupt the star. This create a series of energetic pulses which eject material from the star before it collapses into a BH. PISNe cause massive stars with He-core masses between approximately 60 and $150 \, \mathrm{M}_\odot$ to be completely disrupted. Therefore, PISNe put a second theoretical mass gap into the distribution of BH masses
    \citep{1964ApJS....9..201F,1967ApJ...148..803R,1967PhRvL..18..379B}. On the other hand, PPISNe affect pre-supernovae stars with He-core masses between around 35 and $60 \, \mathrm{M}_\odot$ enhancing the loss of mass before the supernova event and resulting in less massive BHs \citep{2016MNRAS.457..351Y,2017ApJ...836..244W,2019ApJ...882...36M}. Neglecting these two phenomena leads us to overestimate the mass of the most massive BHs with $M_\mathrm{BH} \gtrsim 35 \, M_\odot$. For a recent population synthesis study of the effect of PISNe and PPISNe on the population of coalescing BBHs using the same code as in this work see \citet{2019ApJ...882..121S}.
    
    During a supernova, the asymmetric ejection of matter \citep[][]{1994A&A...290..496J,1996PhRvL..76..352B,2013MNRAS.434.1355J} or asymmetric emission of neutrinos \citep[][]{1993A&AT....3..287B,2005ApJ...632..531S} can provide a momentum kick to the newly formed compact object. Here we assume that the birth kicks of BHs follow a Maxwellian distribution with \mbox{$\sigma = 265$ km/s} \citep{2005MNRAS.360..974H}, which is then rescaled by one minus the fall-back mass fraction $f_{fb}$ \citep{2012ApJ...749...91F}. In the \citet{2012ApJ...749...91F} that we adopt, this quantity depends on the carbon-oxygen core mass $m_\mathrm{core}$ of the star before the collapse. For core masses grater than $11 \, \text{M}_\odot$, $f_{fb} = 1$, which means that in our model all heavy black holes receive no natal kicks. These kicks can tilt the orbit of the BBH, which may generate a negative $\chi_\mathrm{eff}$, add eccentricity to the orbit or disrupt the binary. We take into account all these orbital changes, as well as orbital changes due to symmetric mass loss, following the analytical calculations of \citet{1996ApJ...471..352K} and \citet{2017ApJ...850L..40A}. 
    
   \begin{figure}
   \centering
   \includegraphics[width=8.5cm]{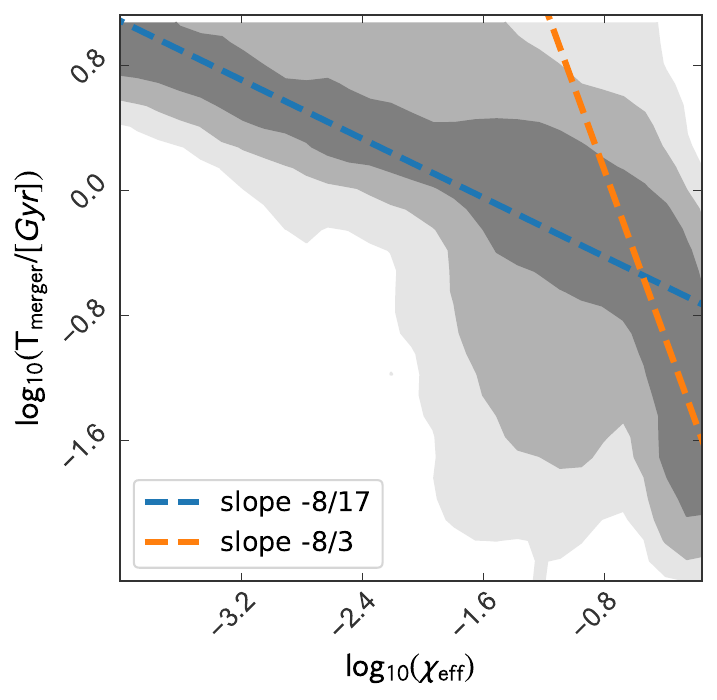}
   \caption{Combined distribution of the BBH merger timescale $\mathrm{T}_\mathrm{merger}$ versus the effective inspiral spin parameter $\chi_\mathrm{eff}$ for our synthetic BBH population at metallicity $\log_{10}(Z) = -2.5$, in gray. The lighter shades represent larger contour levels, 68, 95 and 99\%, respectively. Tidally locked systems follow the relation $\log_{10} (\mathrm{T}_\mathrm{merger}) \sim - \frac{8}{3} \log_{10}(\chi_\mathrm{eff})$, orange dashed line, while the others systems follow $\log_{10} (\mathrm{T}_\mathrm{merger}) \sim - \frac{8}{17} \log_{10}(\chi_\mathrm{eff})$ dictated by the tidal synchronization timescale, blue dashed line. Both lines are drawn at an arbitrary ordinate.}
    \label{fig:tides}
    \end{figure}

\textbf{\subsection{Inspiral due to gravitational waves}\label{sec:inspiral}} 

    After the birth of the second-born BH, GW emission removes energy and angular momentum from the orbit, shrinking it, and eventually leading to the merger of the two compact objects. The merger timescale for eccentric BBHs is computed as
    \begin{equation}
        \mathrm{T}_{\rm merger} = \frac{15}{304} \frac{c^5}{G^3} \frac{1}{m_1m_2(m_1+m_2)} A^4 f(e) \, ,
    \end{equation}
    where $m_1$ and $m_2$ are the masses of the BHs, $A$ is the orbital separation \citep{PhysRev.136.B1224} and $f(e)$ is a numerical factor that account for the orbital eccentricity:
    \begin{equation}
        f(e) = \frac{ \left(1-e^2\right)^4 \int_0^e \frac{e'^{29/19}(1+\frac{121}{304}e'^2)^{1181/2299}}{(1-e'^2)^{3/2}} de' }{e^{48/19} (1+\frac{121}{304}e^2)^{3480/2299}} 
         \, .
    \end{equation}
    
    There is an important point to make here. As was already explained by other authors \citep[e.g.,][]{2016MNRAS.462..844K,2018MNRAS.473.4174Z,2018A&A...616A..28Q}, the merger timescale is anti-correlated with the spin of the second-born BH, $a_2$, or the observed quantity $\chi_\mathrm{eff}$. This is because in order to form a fast rotating BH, tides should be strong and therefore the orbital separation between the parent He-star and the BH companion should be small. Since the merger timescale scales as the fourth power of the orbital separation, for tidally locked systems we can recover the following proportionality $T_\mathrm{merger} \sim A^4 \sim \omega^{-8/3} \sim a_2^{-8/3} \sim \chi_\mathrm{eff}^{-8/3}$. In the second relation we used Kepler's third law with $\omega$ being the orbital frequency matching the He-star's angular frequency $\Omega$ and in the last one $a_1 = 0$ as assumed in our model. Meanwhile, for the wider binaries partially synchronized by dynamical tides on a synchronization timescale $T_\mathrm{sync}^{-1} = |\dot{\Omega}|/\Omega$, we recover small spins $a_2 \sim \Omega \sim \exp(1/T_\mathrm{sync}) \sim 1/T_\mathrm{sync} \sim A^{-17/2}$ (cf. Eq. (\ref{eq:tides})) and therefore $T_\mathrm{merger} \sim A^4 \sim a_2^{-8/17} \sim \chi_\mathrm{eff}^{-8/17}$. 
    In Fig.~\ref{fig:tides} we show the combined distribution of the merger timescale and the effective spin parameter for a specific metallicity bin (centered at $\log_{10}(Z) = -2.5$) of our synthetic BBH population, namely not integrating over redshift and not accounting for any selection effect, in gray. Indeed, systems with high $\chi_\mathrm{eff}$ follow the scaling relation for tidally locked systems (orange dashed line), while those with low $\chi_\mathrm{eff}$ follow the scaling dictated by the tidal synchronization timescale (blue dashed line).

   \begin{figure*}
   \centering
   \includegraphics[width=15cm]{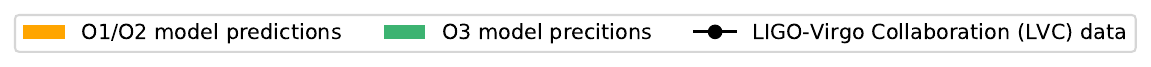}
    \begin{minipage}[b]{0.33\textwidth}
    \includegraphics[width=\textwidth]{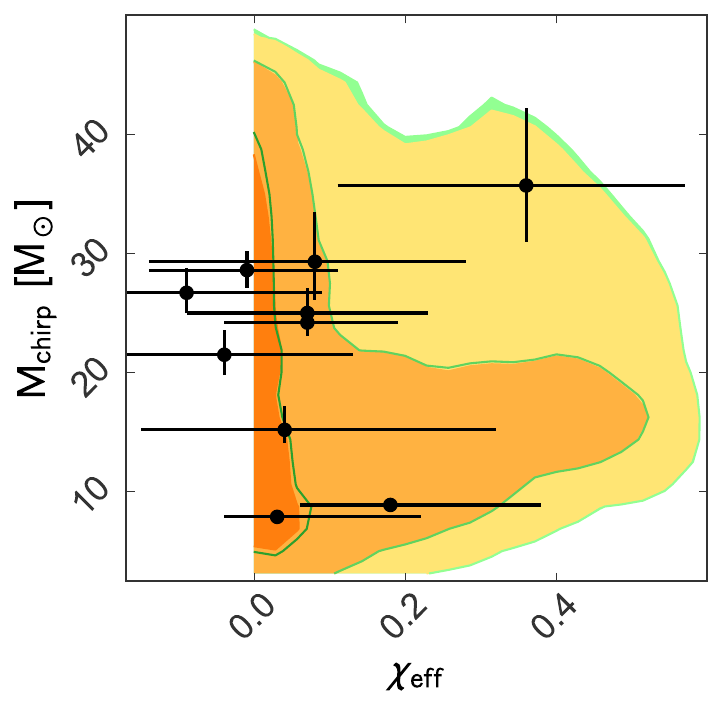}
    \end{minipage}
    \begin{minipage}[b]{0.33\textwidth}
    \includegraphics[width=\textwidth]{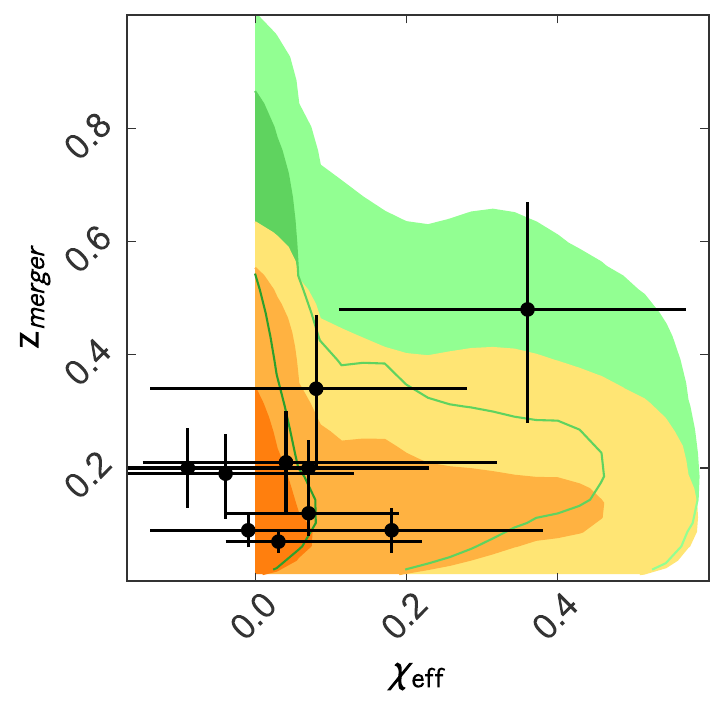}
    \end{minipage}
    \begin{minipage}[b]{0.33\textwidth}
    \includegraphics[width=\textwidth]{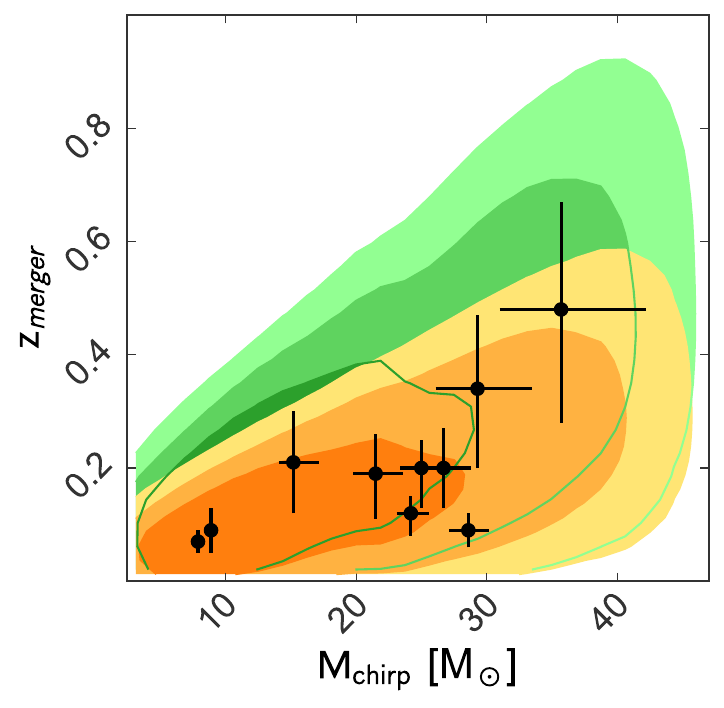}
    \end{minipage}
   \caption{Model predictions for binary black hole observables: chirp mass $\mathrm{M}_\mathrm{chirp}$, effective inspiral spin parameter $\chi_\mathrm{eff}$ and cosmological redshift of merger $z_\mathrm{merger}$ distributions. We represent O1/O2 observing runs in orange and O3 in green, while lighter colors represent larger contour levels of 68, 95 and 99\%, respectively. We overlaid in black the O1/O2 LIGO-Virgo Collaboration (LVC) data with their 90\% credible intervals.}
    \label{fig:ModelPredictionsO}
    \end{figure*}

\textbf{\subsection{Detection rate}\label{sec:R_det}}
    To compute the expected rate of detectable GW events, we need to convolve the star-formation rate (SFR) and metallicity distribution at different redshift epochs with the selection effects of the detectors. To do this we follow a similar approach to the one used in \citet{2016ApJ...819..108B}. Here we briefly summarize our approach, which is described in detail in Appendix \ref{app:DetectionRate}.
    
    In our cosmological calculation we adopt the flat $\Lambda$CDM model with $H_0=67.7 \, \frac{\text{km/s}}{\text{Mpc}}$ and $\Omega_m = 0.307$ \citep{2016A&A...594A..13P}. Every simulated BBH $k$ with BH masses $m_{1,k}$ and $m_{2,k}$, born at redshift $z_{f,i}$ and merging at redshift $z_{m,i,k}$ set by the delay time of this binary contributes to the detection rate by the following weight
    \begin{equation}
        w_{i,j,k} = \frac{\text{fSFR}(z_{f,i})}{M_{\mathrm{ sim},\Delta Z_j}} \, f_{\rm corr} \, 4 \pi c\, D_c^2(z_{m,i,k})\, p_{\rm det}(z_{m,i,k}, m_{1,k}, m_{2,k}) \, \Delta t_i 
        \label{eq:weights}
    \end{equation}
    where subscripts $f$ and $m$ refer to formation and merger, respectively, and fSFR is the fractional star-formation rate, that is the total mass of stars formed per comoving volume per year per metallicity interval $\Delta Z_j$. We adopt the SFR and metallicity distribution of \citet{2017ApJ...840...39M}. The SFR formula we adopt is computed from UV and infrared surveys and is an updated version of \citet{2014ARA&A..52..415M} that better reproduces recent results at high redshifts $4 \leq z \leq 10$ \citep[][]{2015MNRAS.452.1817B,2015ApJ...810...71F,2015ApJ...799...12I,2015MNRAS.450.3032M,2015ApJ...808..104O,2016MNRAS.459.3812M}. The metallicities are log-normally distributed with standard deviation 0.5 dex around the mean metallicity function of \citet{2017ApJ...840...39M}. The mean metallicity function is obtained fitting observations assuming that the galaxy mass -- metallicity relation holds at any redshift \citep{2009ApJ...702.1393K,2012MNRAS.421..621B,2012ApJ...752...66L,2013A&A...556A..55I,2015A&A...575A..96G}. Furthermore, $M_{\mathrm{sim},\Delta Z_j}/f_{\rm corr}$ is the matter simulated in the metallicity bin $\Delta Z_j$ rescaled by the normalization factor $f_\mathrm{corr}$ (see Appendix \ref{app:Normalization}), $D_c(z) = c/H_0 \int_0^z (\Omega_m (1+z')^3+\Omega_\Lambda)^{-1/2} dz'$ is the comoving distance to the source and $p_\mathrm{det}$ accounts for the selection effects of the detector. The total rate of detectable BBH mergers for a given detector network is calculated from the Monte Carlo simulations as a sum
    \begin{equation}
       R_{\rm det} = \sum_{\Delta t_i} \sum_{\Delta Z_j} \sum_k  w_{i,j,k} \, ,
       \label{eq:Rdet}
    \end{equation}
    where we add the contribution of every binary placed at the center of each formation time bin $\Delta t_i$ in its corresponding metallicity bin $\Delta Z_j$. The population synthesis predictions are performed in finite time bins of $\Delta t = 100$ Myr and the log-metallicity range $Z \in [0.0001,0.0349]$ is divided in 30 bins. 
    
    The sensitivity of a GW detector to a source depends on the distance to the source, its orientation and position relative to the detector and on its physical characteristics. The detectability of a signal depends on its signal-to-noise ratio (S/R). In our model we assume that signals are detected if their single-detector S/R exceeds a threshold value of 8 \citep[][]{2016PhRvD..93d2006A}. For the two observing runs O1/O2 of aLIGO, we assumed a detector sensitivity equal to the target ``early high sensitivity'' \citep{2018LRR....21....3A}. This simplification is motivated by the fact that the sensitivity of O2 was close to that of O1 \citep{2018LRR....21....3A}. We assume a target ``late low sensitivity'' for the third observing run O3 and for design sensitivity the corresponding one \citep{2018LRR....21....3A}. We follow the methodology and implementation of \citet{2018MNRAS.477.4685B} [see their section 3.2] to compute the detection probability $p_\mathrm{det}$ for a given set of parameters $(m_1,m_2,z)$.  The optimal S/R (for a face-on, i.e. zero inclination, directly overhead source) is computed for a single detector using the sensitivity above with GW waveforms from \texttt{lalsuite} \citep[][]{lalsuite}. This S/R is then convolved with the antenna pattern function distribution \citep{PhysRevD.47.2198} in order to efficiently sample over the four angles involved, two for the sky location and two for the source orientation, which allows us to estimate the probability of detection. In our simplification we ignored the impact of BH spin on detectability, although high $\chi_\mathrm{eff}$  may slightly enhance $p_\mathrm{det}$.
        
    \begin{figure*}
        \centering
        \begin{minipage}[b]{0.49\textwidth}
        \includegraphics[width=\textwidth]{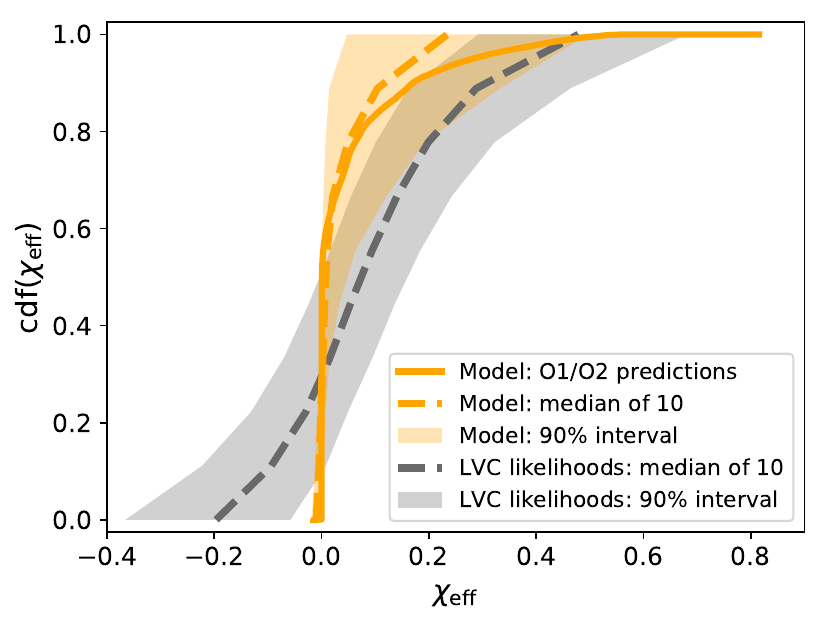}
        \end{minipage}
        \begin{minipage}[b]{0.49\textwidth}
        \includegraphics[width=\textwidth]{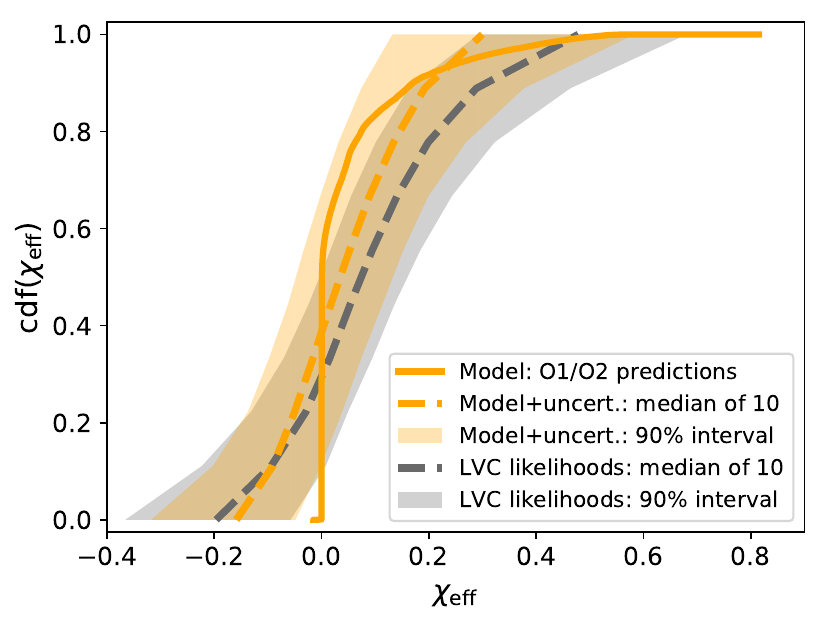}
        \end{minipage}
        \caption{Cumulative distribution functions (cdfs) of the effective inspiral spin parameter $\chi_\mathrm{eff}$ as predicted for O1/O2 by our model (solid orange line). (\textit{Left}) From our model we generate 5,000 sets of 10 measurements and plot the median cdf (dashed orange line) together with the 90\% credible interval (shaded orange region). (\textit{Right}) We generate again 5,000 sets of 10 measurements from our model to which we now add mock measurement uncertainties generated from the zero-centered LVC data likelihoods. We plot the median cdf (dashed orange line) and the 90\% credible interval (shaded orange region). For both graphs, in gray, we plot the median cdf (dashed line) and the 90\% credible interval (shaded region) of 5,000 samples from the 10 actual observations generated from their respective LVC data likelihoods.}
        \label{fig:cdfchi}
    \end{figure*}

\section{Results}\label{sec:results}
    
    We use our model to predict the distributions of the three main observables inferred from GW detections: the chirp mass, the effective inspiral spin parameter and the cosmological merger redshift \citep{PhysRevLett.116.241102}. Every binary in our population synthesis model contributes to the total distributions of every observable quantity with the weight given in Eq. (\ref{eq:weights}).
    
    Our detailed binary evolution models give predictions about the spin of the second-born BH and its misalignment with the orbit. However, in order to estimate the observable $\chi_{eff}$, we need to also have information about the spin of the first-born BH. As we already discussed earlier, here we assume that the spin of the first-born BH is very low, $a_1 \simeq 0$. This is due to two reasons. First, while the progenitor of the BH evolves through the red supergiant phase, most of the angular momentum is transported to the outer layers of the star upon expansion (because of the assumed efficient AM transport). This depletes the angular momentum of the core and eventually, due to mass transfer or stellar winds which remove the outer layers of the star, leads to a slowly rotating naked He-star. Second, the initial orbital separation of the two stars is quite large compared to the later stage of the evolution. Thus, even if tides can efficiently synchronize the rotation of the star to the orbit, the angular frequency of the envelope is too low to efficiently spin up the core. To quantitatively check these assumptions, \citet{2018A&A...616A..28Q} used detailed stellar-evolution simulations to show that main sequence stars with initial angular rotations up to $\Omega_\mathrm{initial} \lesssim 0.5 \, \Omega_\mathrm{critical}$ evolve to yield BHs with negligible spins at all metallicities, even when assuming that  the angular momentum of the core is conserved in the collapse.  A small subset of the most rapidly spinning stars undergo efficient internal mixing and evolve chemically homogeneously. These stars never expand to become giant stars and hence do not evolve through the standard CE formation channel.

\textbf{\subsection{aLIGO O1, O2, \& O3 observing runs}}

    The first and second observing runs of aLIGO (and, for parts of it, Advanced Virgo) lasted for 4 and 9 months, respectively, resulting in a total of 166 days of data suitable for coincident analysis \citep[][]{2016PhRvX...6d1015A}. Ten GW signals from BBH mergers were detected \mbox{\citep[][]{2019PhRvX...9c1040A}}. These ten detections translate to a rate of 22 BBH mergers per year. In our model comparison to the data we only include these ten detections from the LIGO-Virgo Collaboration's catalog, although evidence for additional signals in O1/O2 data has been presented by \citet{2019arXiv191009528Z,2019PhRvD.100b3007Z,2019arXiv190407214V}.
    Our goal here is to model the combined distributions of observable quantities of the CE formation channel, with a special focus on the spins of the BHs which we investigate in detail.  We intentionally picked a population synthesis model that approximately matches the observed rate for this study. Using Eq. (\ref{eq:Rdet}), our model predicts for O1/O2 a detection rate of 27 yr$^{-1}$, while for the ongoing observing run O3, we predict a detection rate of around \mbox{100 yr$^{-1}$}.  However, the predicted event rate depends sensitively on a number of uncertain evolutionary model parameters \citep[e.g.,][]{2015ApJ...806..263D,2018MNRAS.474.2959G,2018MNRAS.477.4685B} and metallicity-specific star formation history \citep[e.g.,][]{2019MNRAS.482.5012C,2019MNRAS.490.3740N}.

    In Fig.~\ref{fig:ModelPredictionsO} we show the predicted two-dimensional distributions of chirp mass, effective inspiral spin parameter and cosmological merger redshift for O1/O2 in orange and O3 in green. Lighter colors delineate larger contour levels of 68, 95 and 99\%, respectively. For a comparison with the observations, we overlay the ten GW detections with their 90\% credible intervals in black. These detections agree visually very well with our model prediction. In the first histogram, $\mathrm{M}_\mathrm{chirp}$ vs. $\chi_\mathrm{eff}$, we see that the selection effects of the detectors at different sensitivities do not significantly affect the chirp mass and the effective inspiral spin parameter distributions. Meanwhile in the other two histograms, $z_\mathrm{merger}$ vs. $\chi_\mathrm{eff}$ and $z_\mathrm{merger}$ vs. $\mathrm{M}_\mathrm{chirp}$, we see that at higher detector sensitivity we are able to detect events at higher cosmological redshift, namely at further distances, and that more massive sources can be observed out to a higher redshift, as one might trivially expect.

\begin{figure}
        \centering
        \includegraphics[width=9cm]{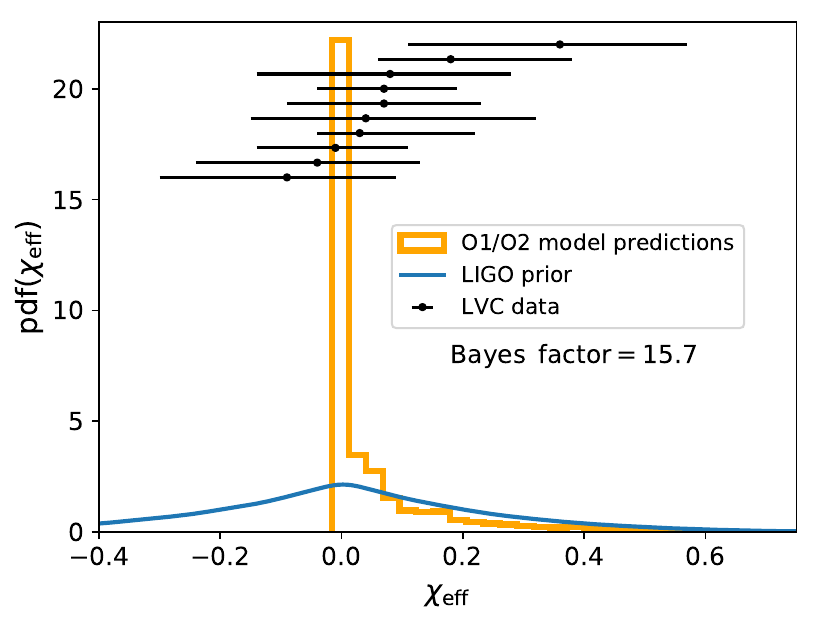}
        \caption{Probability density function of $\chi_\mathrm{eff}$ predicted by our O1/O2 model (orange) and the average LIGO prior (blue). At an arbitrary vertical position, we plot in black the LVC data with their respective 90\% credible intervals. The Bayes factor between our model and the LIGO prior is 15.7.}
        \label{fig:pdfchi}
    \end{figure}

    To provide a qualitative comparison between our theoretical predictions of $\chi_\mathrm{eff}$ and LIGO-Virgo Collaboration (LVC) measurements, we conducted a visual cumulative distribution function (cdf) graphical check and a Bayesian model comparison between our model and the LVC prior (uniform spin magnitudes $a_{1,2} \in [0,1]$ and isotropic spin directions).
    For the graphical check, shown in Fig.~\ref{fig:cdfchi}, we generate 5,000 sets of 10 mock events from a KDE of our O1/O2 model predictions (solid orange line in Fig.~\ref{fig:cdfchi}) and compute the corresponding cdf. In the left panel of Fig.~\ref{fig:cdfchi} we plot the median cdf (dashed orange line) and 90\% credible interval (shaded orange region) of these sets of mock observations without any measurement uncertainty. Our model mostly predicts positive $\chi_\mathrm{eff}$ and only a few slightly negative $\chi_\mathrm{eff}$ but cannot reproduce $\chi_\mathrm{eff} \ll 0$. In the right panel of Fig.~\ref{fig:cdfchi} we plot the same quantities, but this time we add uncertainties to the 5,000 sets generated from our model. These uncertainties are generated from the zero-centered LVC likelihoods.
    We show the cdfs from 5,000 sets of 10 samples, one each from the LVC data likelihoods of the 10 observed events, in gray in both panels of Fig.~\ref{fig:cdfchi}. These likelihoods are obtained by weighting the posteriors of the ten O1/O2 GW observations by the inverse of the average projected LVC prior, which is found by combining the samples of all ten priors. We see that, when accounting for observational uncertainties, we can also reproduce negative values of $\chi_\mathrm{eff}$ as in the observed cdf tail. We conclude that, visually, our model agrees well with the data.
    Of course, there is no single statistical check to unambiguously test the goodness of a model.  In addition to the graphical check described above, we could ask, for example, whether the existing observations are statistically consistent with a model that predicts a negligible number of events with negative $\chi_\mathrm{eff}$.  They are indeed consistent.  All individual observations allow for zero or positive values of $\chi_\mathrm{eff}$ well within their 90\% credible intervals.   However, if future GW observations have significant negative $\chi_\mathrm{eff}$ inconsistent with zero, this will be an indication that those systems have been formed through alternative channels, such as dynamical formation.
    In Fig.~\ref{fig:pdfchi} we show the probability density function of $\chi_\mathrm{eff}$ as predicted by our O1/O2 model, in orange, and the average projection of the LVC prior onto $\chi_\mathrm{eff}$. For reference we added the ten LVC GW observations with their 90\% credible intervals at an arbitrary vertical position. We carried out a Bayesian model comparison between our model and the LVC prior given the observational data. This test results in a Bayes factor of 15.7 which favors our model.

    \begin{figure*}
        \centering
        \includegraphics[width=18cm]{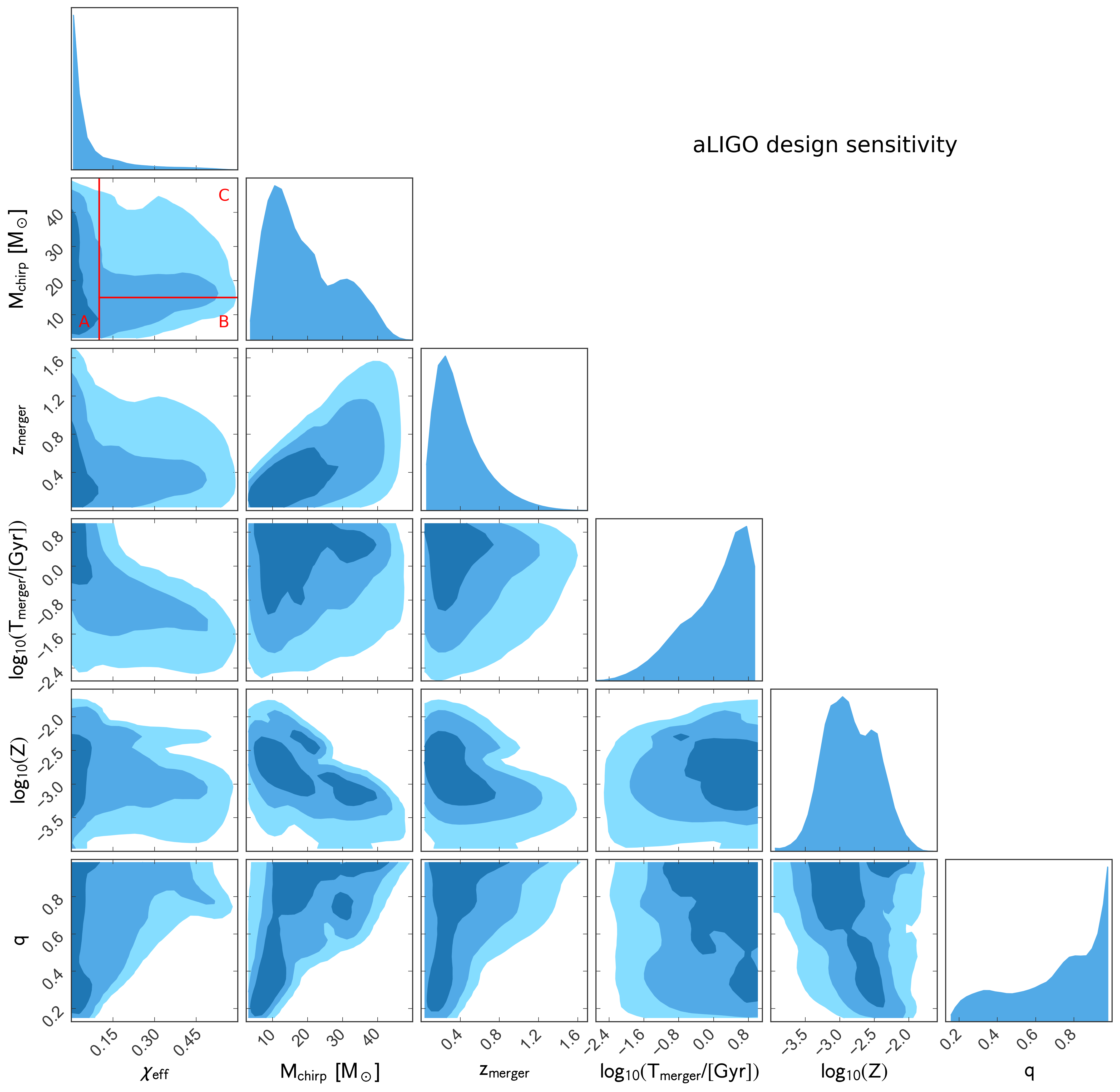}
        \caption{Model predictions for the chirp mass $\mathrm{M}_\mathrm{chirp}$, effective inspiral spin parameter $\chi_\mathrm{eff}$, cosmological redshift of merger $z_\mathrm{merger}$, BBH merger timescale $\mathrm{T}_\mathrm{merger}$, metallicity $Z$ and the binary mass ratio $q$ distributions of binary black holes observables at design sensitivity. Lighter colors represent larger contour levels of 68, 95 and 99\%. We arbitrarily divide the two dimensional histogram $\mathrm{M}_\mathrm{chirp}$ vs $\chi_\mathrm{eff}$ with red lines into three regions at $\chi_\mathrm{eff} = 0.1$ and $\mathrm{M}_\mathrm{chirp} = 15 \, \mathrm{M}_\odot$. Region-A contains 80\% of the events, 10\% are in Region-B and 10\% are in Region-C. For illustrative purposes, all histograms are plotted with a smoothing scale of 0.8 bins with the exception of $z_\mathrm{formation}$ vs. $\chi_\mathrm{eff}$, $z_\mathrm{formation}$ vs. $z_\mathrm{merger}$, $q$ vs. $\chi_\mathrm{eff}$ that have no smoothing}.
        \label{fig:Design}
    \end{figure*}

    \begin{figure*}
        \centering
        \includegraphics[width=18cm]{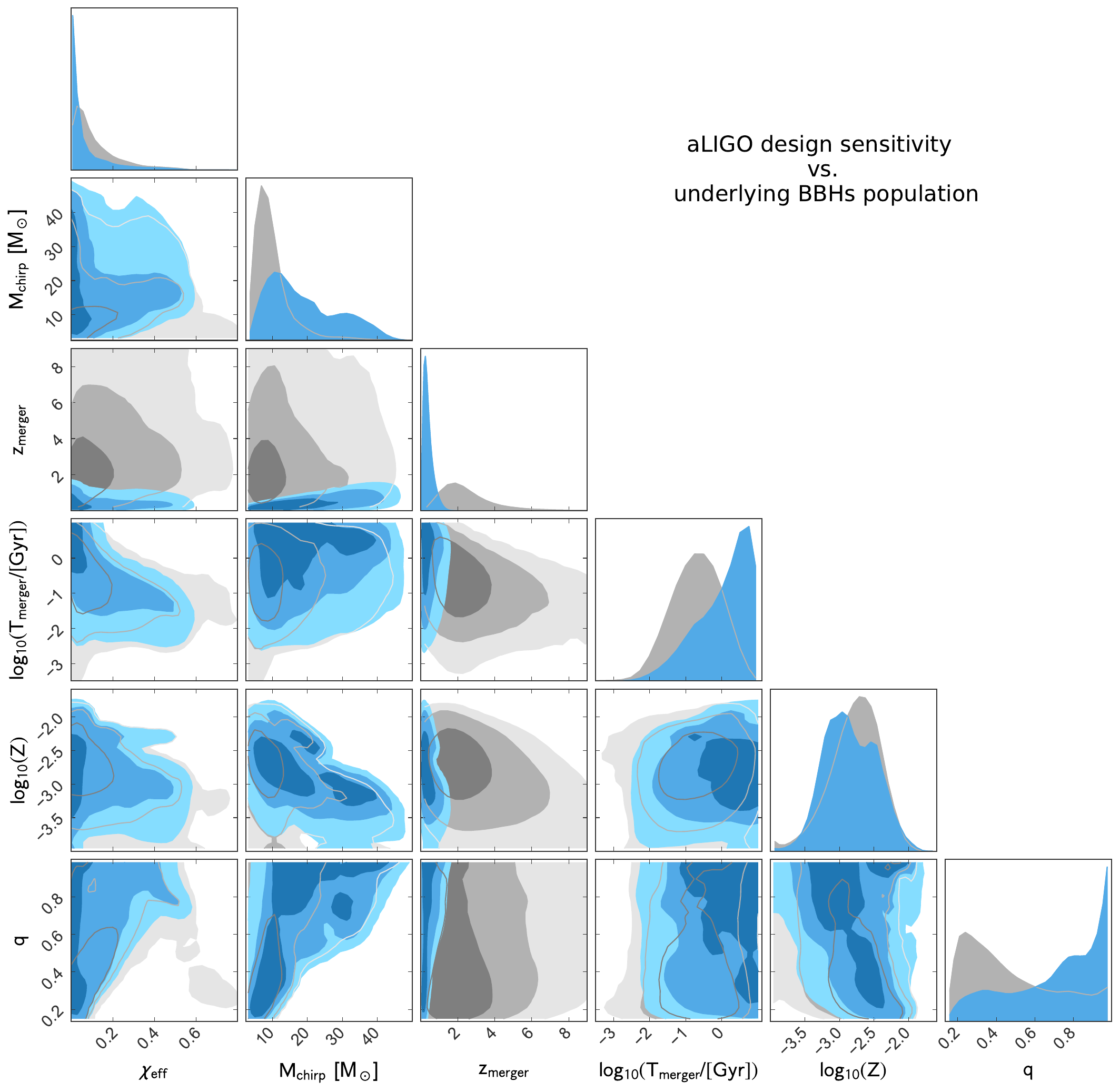}
        \caption{Model predictions for the chirp mass $\mathrm{M}_\mathrm{chirp}$, effective inspiral spin parameter $\chi_\mathrm{eff}$, cosmological redshift of merger $z_\mathrm{merger}$, BBH merger timescale $\mathrm{T}_\mathrm{merger}$, metallicity $Z$ and the binary mass ratio $q$ distributions of binary black holes observables at design sensitivity, in blue, versus the underlying population of merging BBHs one would observe with a GW detector with infinite sensitivity, in gray. Lighter shades represent larger contour levels of 68, 95 and 99\%, respectively.}
        \label{fig:DesignVsWhole}
    \end{figure*}

\textbf{\subsection{aLIGO design sensitivity}}

    We use the target design sensitivity curve of \citet{2018LRR....21....3A} and predict a BBH merger rate of around 400 yr$^{-1}$ for advanced detectors operating at design sensitivity. In Fig.~\ref{fig:Design}, we show the expected properties of the BBH population detectable at aLIGO design sensitivity: the effective inspiral spin parameter, chirp mass and cosmological merger redshift, as well as the BBH merger timescale, metallicity and binary mass ratio. 
    
    As pointed out previously, the two-dimensional distribution of the effective inspiral spin parameter vs. the chirp mass does not change with different detector sensitivities. We arbitrarily divide the parameter space and identify three regions: Region-A with $\chi_\mathrm{eff}<0.1$, Region-B with $\chi_\mathrm{eff} \ge 0.1$ and $\mathrm{M}_\mathrm{chirp} < 15 \, \text{M}_\odot$ and Region-C with $\chi_\mathrm{eff} \ge 0.1$ and $\mathrm{M}_\mathrm{chirp} \ge 15 \, \text{M}_\odot$. Our model predicts that 80, 10 and 10\% of detectable BBH mergers fall into these three regions, respectively.

    To understand these different regions of the parameter space, we recall that there is an anti-correlation between the merging timescale of the BBHs $T_{\rm merger}$ and the spin of the second-born BH $a_2$ or equivalently the observed quantity $\chi_\mathrm{eff}$ (see Sec. \ref{sec:inspiral}). Region-A contains systems with low spins which translate into long merger timescales. For reference, at formation of the second-born BH most of the orbital periods are between 1 and 5 days. These systems might have formed at redshifts up to 10 and they probe a wide range of metallicities and chirp masses. BBHs in Region-B have short merger timescales since they have a high $\chi_\mathrm{eff}$: at the formation of the second-born BH they are in close orbits with periods smaller than 1 day. These BBHs are formed in the local Universe, at redshifts close to zero, where the average metallicity is high and the chirp mass is low because high metallicity massive stars tend to lose a lot of mass due to stellar winds and thus the resulting BH masses are lower. Finally, systems in Region-C have high spins and high chirp masses. These are binaries that formed with low metallicity and merge quickly, that is at $\mathrm{z}_\mathrm{merger} \simeq \mathrm{z}_\mathrm{formation}$. This part of the parameter space really probes the low-end tail of the metallicity distribution out to the observational redshift horizon of aLIGO. These are intrinsically rare systems but are amplified by aLIGO's higher sensitivity to high BH masses.

    In Fig.~\ref{fig:DesignVsWhole} we further investigate GW selection effects that favor high BH masses. We show, in blue, the model prediction for aLIGO at design sensitivity against the overall BBH underlying population that one would observe with an infinitely-sensitive detector, in gray. The entire population of merging BBHs has a peak in the merger redshift at around $\mathrm{z}_\mathrm{merger} \simeq 2$.  While aLIGO is not sensitive to mergers at such high redshifts, future third generation GW detectors, such as the Einstein Telescope \citep[][]{2010CQGra..27s4002P,Kalogera:2019}, are able to observe them. The selection effects in favor of higher BH masses are clearly visible in the distribution of $\mathrm{M}_\mathrm{chirp}$. Our observable predictions show a bimodal distribution of chirp masses with peaks at 11 M$_\odot$ and 33 M$_\odot$, while the underlying population has only one peak at around \mbox{10 M$_\odot$}. Selection effects allow us to observe massive BHs formed at high redshifts where the mean metallicity is lower than today. GW detectors preferentially observe BBHs that do not merge quickly, namely have wider orbits and slower spin (see the peak at $\chi_\mathrm{eff} \simeq 0$ in the $\chi_\mathrm{eff}$ histogram). We note that in our treatment of the selection effects we did not take into account the potentially greater sensitivity to GWs from BBHs with high $\chi_\mathrm{eff}$; this may influence the tail of the effective inspiral spin parameter distribution, accentuating the second peak at $\chi_\mathrm{eff}=0.4$.

    \begin{figure}
        \centering
        \includegraphics[width=9cm]{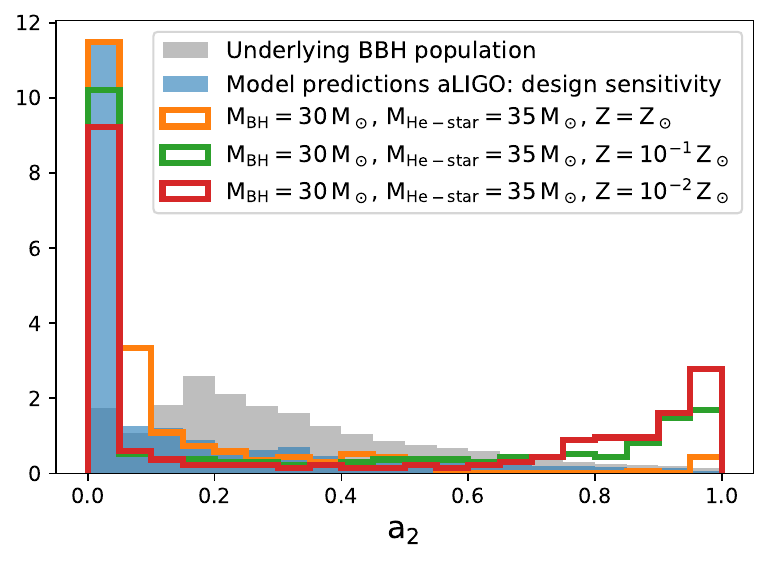}
        \caption{Normalized distribution of the spin of the second-born BH as predicted by our model for GWs observable at aLIGO design sensitivity, in light blue, versus the underlying population of merging BBHs, in gray. For comparison, we show the same distribution obtained with MESA simulations of BH-He-star binary systems with $\mathrm{M}_\mathrm{BH}=30$ M$_\odot$, $\mathrm{M}_\mathrm{He-star}=35$ M$_\odot$ and metallicities  $\mathrm{Z}_\odot$ (orange), $10^{-1} \, \mathrm{Z}_\odot$ (green) and $10^{-2} \, \mathrm{Z}_\odot$ (red) under the assumption that the initial separation is log-uniform in $ 5 < A / \mathrm{R}_\odot < 63$.}
        \label{fig:a2}
    \end{figure}

\vspace{0.5cm}

\section{Discussion}\label{sec:discussion}

\textbf{\subsection{Angular momentum efficiency}}

    Our results are obtained assuming efficient angular momentum transport \citep[][]{1999A&A...349..189S,2002A&A...381..923S,2019MNRAS.485.3661F} which plays an important role in determining the spin of the first-born BH.  Meanwhile, the spin of the second-born BH is mainly determined by the combined effects of stellar wind and tidal interaction during the binary evolution. Although the Tayler-Spruit dynamo model helps to reproduce the flat rotation profile of our Sun \citep[][]{2014ApJ...796...17F,2014ApJ...788...93C} as well as neutron star and white dwarf spins \citep[][]{2005ApJ...626..350H,2008A&A...481L..87S}, it cannot reproduce the asteroseismic constrains for sub-giants and red giants \cite[][]{2018A&A...616A..24G}, which would require an even higher efficiency in angular momentum transport. Alternatively, a model with inefficient angular momentum transport predicts highly spinning BHs, $\chi_\mathrm{eff} \simeq 1$ \citep[][]{2017arXiv170607053B,2019MNRAS.482.2991A}, which do not match current GW observations. To test that angular momentum transport efficiency affects only the spin of the first born BH, and perhaps the initial rotation of the helium star  after the common envelope, but not how tides can spin up an initially non-spinning helium star, we ran a grid of 5,000 MESA simulations of BH-He-star binaries with inefficient angular momentum transport, namely without the Tayler-Spruit dynamo. We found that there is a negligible impact on the spin of the second-born BH. 
    
\textbf{\subsection{Comparison with other studies}}

    When comparing our results with other studies of the CE channel, we find some discrepancies. For example \citet{2018MNRAS.473.4174Z} found a bimodal distribution of spins of the second-born BH, with around half of the BHs having spin zero and the other half maximally spinning. When accounting for stellar winds and tidal interaction between the BH and He-star in a detailed binary evolution calculation we find that this bimodal distribution is an oversimplification. In Fig.~\ref{fig:a2} we show the normalized distribution of the spin of the second-born BH $a_2$ from detailed BH-He-star binaries simulations with masses \mbox{$\mathrm{M}_\mathrm{BH} = 30$ M$_\odot$}, $\mathrm{M}_\mathrm{He-star}=35$ M$_\odot$ and metallicities $Z_\odot, 10^{-1} Z_\odot, 10^{-2} Z_\odot$ assuming a uniform distribution in log-orbital separation. Indeed, we find an approximately bimodal distribution of spins (similar to Fig. 4 from \citet{2018MNRAS.473.4174Z}) at low metallicity where stellar winds are weak and do not affect the orbital evolution. However, at higher metallicity the wind-driven mass loss causes the binaries to widen and the tidal interaction which spins up the He-star gradually becomes less effective. 
    In the same figure we also show our predicted $a_2$ distribution from our model of BBH mergers observable at the aLIGO design sensitivity (blue shaded region) versus the underlying population of merging BBHs (gray shaded region). The former one has a peak at around $a_2=0$ while the latter has a peak at around $a_2 = 0.2$. Although it has a peak at $a_2=0$, it is much flatter and both do not show the second peak at high spins. A key reason is that we did not assume a log-normal distribution of orbital separation after the CE phase, as was assumed by \citet{2018MNRAS.473.4174Z}, but used the predictions of our population synthesis model (cf.~Fig \ref{fig:postCE}). Moreover, we also take into account the redshift- and metallicity-dependent star formation rate and apply aLIGO selection effects. The anti-correlation between $T_\mathrm{merger}$ and $a_2$ means that binary systems with high spins merge quickly. Thus, their merger redshift distribution follows the SFR evolution with a peak around $z\simeq 2$, beyond aLIGO sensitivity. This further reduces the number of rapidly spinning BHs detectable by aLIGO. We believe that the non-inclusion of detailed binary evolution calculations that carefully track the angular momentum evolution due to tidal interaction and stellar winds is also the reason for overestimated $\chi_\mathrm{eff}$ distributions derived in other studies \citep{2018PhRvD..98h4036G,2019MNRAS.483.3288P}.

\textbf{\subsection{Comparison with other formation scenarios}}

    We now compare our findings with theoretical results from other formation scenarios. All isolated binary evolution channels produce BH spins which are expected to be mostly aligned. It was shown by \citet{2016A&A...588A..50M} that the chemically homogeneous evolution channel mostly produce highly spinning BHs ($a_{1,2} > 0.4$), which are currently not observed. An indicator that would rule in favor of this scenario are the detections of high effective spin parameters, say $\chi_\mathrm{eff} > 0.8$, which are not predicted by the CE channel. \citet[][]{2019PhRvD.100b3007Z} recently reported the finding of a new BBH merger by reanalyzing the publicly available raw data from O1, using an independently developed pipeline. Assuming a flat $\chi_\mathrm{eff}$ prior they found an event with high chirp mass, \mbox{$\mathrm{M}_\mathrm{chirp}=31^{+2}_{-3}$ M$_\odot$}, and high effective spin parameter, $\chi_\mathrm{eff}=0.81^{+0.15}_{-0.21}$, which is marginally consistent with our model. If similar events with better-measured parameters are found in the future, such BBHs would probably have been formed through the chemically homogeneous evolution channel, since the high $\chi_\mathrm{eff}$ is outside our model prediction range for the CE channel. For dynamically-formed BBHs in dense star cluster, \citet{2018PhRvL.120o1101R} found a symmetric distribution of $\chi_\mathrm{eff}$ with a peak at $\chi_\mathrm{eff}=0$, regardless of the BH birth spins. Therefore, anti-aligned systems ($\chi_\mathrm{eff} < 0$) would be a key indicator of the dynamical formation channel, as these are not predicted for either the CE channel or the chemically homogeneous evolution channel.

\textbf{\subsection{Uncertainties}}

    Our model may be limited by some uncertainties which can alter the merger rate and, to a lesser degree, the predicted BBH property distributions. Uncertainties in (i) how the CE phase is accounted for, namely different choices of the $\alpha_{CE}$ parameter which characterizes the efficiency of transferring orbital energy into unbinding the CE might lead to different rate predictions \cite[see e.g.,][]{2018MNRAS.480.2011G}.
    For example, \citet{2019ApJ...883L..45F} report a very efficient (high) $\alpha_{CE}$ for a specific binary neutron stars (BNSs) system analyzed with 1D hydrodynamic simulations. However, in their estimate of $\alpha_{CE}$ they do not include the envelope's thermal energy in the calculation of the envelope's binding energy, and, most importantly, their results may not carry over to BBH formation, which tends to happen at lower metallicities, with more similar donor and accretor masses at common envelope onset. Indeed, \citet{2018MNRAS.479.4391M} showed that uncertainties in $\alpha_{CE}$ correspond to a variation of a factor of around 1.5 in the BBH merger rate estimates while a factor of 10 for BNS merger rates. Another example are uncertainties in the (ii) physics of the supernova explosions, such as the kicks strength, which can influence rates and affect the parameter distribution of BBH mergers \citep{2013ApJ...779...72D}. To test how the delayed SN mechanism affects our results, we relaxed the model to account for direct collapse of the second born BH (still assuming a possible 10\% of mass defect). We found similar distributions to the one of the delayed collapse with only a slight increase around $M_\mathrm{chirp} \simeq 11 \, M_\odot$ and $\chi_\mathrm{eff} \simeq 0.2$ for detectable binaries and an increase of the merger rate by $\sim$ 10\% due to the survival of BBHs disrupted by natal kicks in the \citet{2012ApJ...749...91F} model and the detectability of the slightly more massive BBHs are greater distances.
    Furthermore, in our binary population, (iii) the He-stars after the CE phase are not necessarily zero-age helium main sequence stars, as assumed in the second step of our detailed binary evolution calculations. This is because some of the progenitors of the helium stars overflowed their Roche lobes and entered the common envelope phase after helium burning was initiated in their core. The remaining lifetime of these stars is shorter than the duration of their zero-age helium main sequence. They lose less mass through stellar winds in their remaining lifetimes, which cause the orbits to not widen as much and result in more massive BHs with higher spins. However, we expect that the fraction of stars that enter the CE while burning helium in their core is higher at low metallicities, as low-metallicity stars tend to expand later in their lives. At the same time, stellar winds in these stars are weaker due to the low metallicity, so the overall effect on the population of BBHs is expected to be limited.
    Moreover, our detection rate calculation is affected by (iv) uncertainties in the redshift dependent SFR, (v) redshift-dependent metallicity distribution and (vi) the initial mass function \citep{2015ApJ...814...58D,2019MNRAS.482.5012C,2019MNRAS.490.3740N} which may not be universal \citep[e.g.,][]{10.1046/j.1365-8711.2001.04022.x,Schneider:2018}, but see \citet{FarrMandel:2018}. For example there is an uncertainty in the UV and IR data used to infer the SFR and mean metallicity at high redshift $z > 4$ \citep{2014ARA&A..52..415M}. A SFR favoring lower formation metallicities than the one assumed here would skew our result in favor of systems in Region-C in the $\chi_\mathrm{eff} - M_\mathrm{chirp}$ histogram of Fig.~\ref{fig:Design}, namely BBHs with high $M_\mathrm{chirp}$ and high $\chi_\mathrm{eff}$, at the expense of systems in Region-A, namely BBHs with low $\chi_\mathrm{eff}$.

\vspace{0.5cm}

\section{Conclusions}\label{sec:conclusion}

    One of the biggest open questions in GW astrophysics today is how merging binary black holes are formed. Isolated binaries that go through the CE phase are one of the main proposed formation channels for BBHs. In this work we investigate the combined distributions of masses, spins and merger redshifts of a population of BBHs formed through this channel that would be detectable by advanced GW detectors. We combine binary population synthesis studies with detailed stellar structure and binary evolution simulations. Rapid population synthesis allows us to obtain a population of BBH progenitors: BH-He-star binaries. Meanwhile, the detailed simulations that take into account the effects of differential stellar rotation, tidal interactions, wind mass loss and the evolution of the structure of the He-star, allow us to accurately predict the distribution of the properties of BBH systems at their formation. We then take into account the redshift and metallicity dependence of the star-formation rate together with the selection effects of the detectors to build a model capable of reproducing all observable properties of the current sample of 10 BBH mergers. We also predict what future GW experiments are likely to observe. Our main findings can be summarized in the following points:
    \begin{itemize}
        \item Our model is the first one to use detailed stellar structure and binary evolution simulations to successfully reproduce the observed $\chi_\mathrm{eff}$ population: most with $\chi_\mathrm{eff} \simeq 0$ and a few with positive $\chi_\mathrm{eff}$. Hence, it provides strong support for the CE channel as the dominant formation channel for the observed BBH mergers.
        
        \item We find that the ten O1/O2 GW detections are consistent with having formed through the CE channel. We predict a detection rate of 27 yr$^{-1}$ for a particular set of population-synthesis model assumptions and a specific choice of a metallicity-specific star formation history, which is consistent with the 10 GW detections found in 167 days of total coincident observing time during the first two advanced detector observing runs.

        \item We predict the combined distributions of $\mathrm{M}_\mathrm{chirp}$, $\chi_\mathrm{eff}$ and $z_\mathrm{merger}$ for the current O3 observing run and for future data at design sensitivity. 
        
        \item We distinguish three different regions of observable BBH mergers. At design sensitivity, we expect around 80\% of events with $\chi_\mathrm{eff} < 0.1$ and a wide range of chirp masses: these systems formed in relatively wide orbits (mostly with periods of 1-5 days at the formation of the second-born BH) and might have formed at redshifts up to 10, probing a wide range of metallicities. Around 10\% of events with $\chi_\mathrm{eff} \ge 0.1$ and $\mathrm{M}_\mathrm{chirp}<15 \, \text{M}_\odot$ are BBHs born in close orbits (with orbital periods of less than 1 day at the formation of the second-born BH) in the local Universe at redshift close to 0 where the metallicity is high. These systems merge promptly. Finally, the remaining 10\% of events with $\chi_\mathrm{eff} \ge 0.1$ and $\mathrm{M}_\mathrm{chirp} \ge 15 \, \text{M}_\odot$ are BBHs formed at low metallicity at a range of redshifts; these systems again merge promptly. Efficient spin-up of the secondary, which yields high $\chi_\mathrm{eff}$, requires the BBH to be born in a close orbit, which then allows for a prompt merger through GW emission.
              
        \item We find that the total population of merging BBHs, namely the one that would be observed by a GW detector with infinite sensitivity, has a peak in the merger redshift at around 2, far beyond aLIGO sensitivity.  This peak is set by a combination of the star formation rate, which peaks at a redshift of 2; the metallicity of star formation, which is lower in the early Universe and favors more efficient BBH formation; and the delay time distribution until merger.
        
        \item We show that in order to understand the distribution of BBH spins, population synthesis studies of isolated field binary formation channels should include detailed binary evolution calculations that carefully track the angular momentum evolution due to the tidal interaction and stellar winds which are the origin of the spin of the second-born BH.  Moreover, we find that the assumption of efficient angular momentum transport has a negligible impact on the spin of the second-born BH.
        
    \end{itemize}

\begin{acknowledgements}
      We would like to thank Pablo Marchant, Michael Zevin, Vicky Kalogera and Christopher Berry for helpful suggestions regarding Section 2.4 and Appendix B. This work was supported by the Swiss National Science Foundation Professorship grant (project number PP00P2 176868). This project has received funding from the European Union's Horizon 2020 research and innovation program under the Marie Sklodowska-Curie RISE action, grant agreement No 691164 (ASTROSTAT). This work was performed in part at the Aspen Center for Physics, which is supported by National Science Foundation grant PHY-1607611. The computations were performed in part at the University of Geneva on the Baobab and Lesta computer clusters and at the University of Birmingham. AB is funded by the program Cátedras CONACYT para Jóvenes Investigadores and  by the Danish National Research Foundation (project number DNRF132). SS is supported by the Australian Research Council Centre of Excellence for Gravitational Wave Discovery (OzGrav), through project number CE170100004. All figures were made with the free Python modules Matplotlib \citep{Hunter:2007} and pygtc \citep{Bocquet2016}. This research made use of Astropy,\footnote{http://www.astropy.org} a community-developed core Python package for Astronomy \citep{astropy:2013, astropy:2018}.
\end{acknowledgements}


\bibliographystyle{aa}
\bibliography{aanda}

\onecolumn
\begin{appendix}


\section{Mass renormalization of the population synthesis simulation}\label{app:Normalization}
    We used COMPAS to run a Monte-Carlo simulation and generate a sample of half a million BH - He-star binaries. When we perform binary population synthesis simulations, we only model binary systems, neglecting the population of single stars. Furthermore, to save on computational costs, we restrict the mass of the primary star to a suitable range $m_A < m_\mathrm{primary} < m_B$ so that we only consider initial binaries that can be progenitors of the systems we want to study (this is a basic version of the importance sampling approach described by \citet{Broekgaarden:2019}).  This means that we model only a fraction of the underlying stellar population.  Here we show how to renormalize the population synthesis simulation to the total stellar mass of the underlying stellar population.

    Let us consider a stellar population of total mass $M_*$ with an initial mass function (IMF) of single star masses:
    \begin{equation}
    	f(m) = \left\{ \begin{array}{rl}
               f_0 \left(\frac{m}{m_\mathrm{min}}\right)^{-\alpha_1} & m_\mathrm{min} \leq m \le m_1 \\
               f_0 \left(\frac{m_1}{m_\mathrm{min}}\right)^{-\alpha_1}\left(\frac{m}{m_{1}}\right)^{-\alpha_2} & m_1 \leq m \le m_2 \\
               f_0 \left(\frac{m_1}{m_\mathrm{min}}\right)^{-\alpha_1}\left(\frac{m_2}{m_{1}}\right)^{-\alpha_2}\left(\frac{m}{m_{2}}\right)^{-\alpha_3} & m_2 \leq m \le m_\mathrm{max} 
               \end{array} \right.
    \label{eq:IMF}
    \end{equation} 
    where the constant $f_0$ is defined such that $\int_{m_\mathrm{min}}^{m_\mathrm{max}}f(m) \, dm = 1$. Let $f_\mathrm{bin}$ be the fraction of stars in binaries and assume that the distribution of mass ratios in binaries is flat between 0 and 1, that is $g(q)=1$. Then, the mean mass of a stellar system in the population is
    \begin{equation}
    \begin{split}
        \bar{m}_{\star} &= (1-f_\mathrm{bin})\int_{m_\mathrm{min}}^{m_\mathrm{max}} m f(m)\,dm +               f_\mathrm{bin}\int_{m_\mathrm{min}}^{m_\mathrm{max}}\int_{0}^{1}\left[f(m)g(q)\left(m+qm\right)\right] \, dqdm = \\
        &= f_0 \left(1+\frac{f_\mathrm{bin}}{2}\right)  \left[ \frac{1}{2-\alpha_1} \left(\frac{m_{1}^{2-\alpha_1}-m_\mathrm{min}^{2-\alpha_1}}{m_\mathrm{min}^{-\alpha_1}} \right) + \frac{1}{2-\alpha_2}\left(\frac{m_1}{m_\mathrm{min}}\right)^{-\alpha_1} \left(\frac{m_{2}^{2-\alpha_2}-m_{1}^{2-\alpha_2}}{m_{1}^{-\alpha_2}} \right)  +\frac{1}{2-\alpha_3}\left(\frac{m_1}{m_\mathrm{min}}\right)^{-\alpha_1} \left(\frac{m_2}{m_{1}}\right)^{-\alpha_2} \left(\frac{m_\mathrm{max}^{2-\alpha_3}-m_{2}^{2-\alpha_3}}{m_{2}^{-\alpha_3}} \right)\right] .
    \end{split}
    \end{equation}
    In the case when $m_A, m_B > m_2$, we only model the following fraction of all systems:
    \begin{equation}
        f_{\mathrm{model}} =  f_\mathrm{bin}\int_{m_A}^{m_B}f(m)\,dm = f_\mathrm{bin}f_0\frac{1}{1-\alpha_3}\left(\frac{m_1}{m_\mathrm{min}}\right)^{-\alpha_1} \left(\frac{m_2}{m_{1}}\right)^{-\alpha_2} \left(\frac{m_{B}^{1-\alpha_3}-m_{A}^{1-\alpha_3}}{m_{2}^{-\alpha_3}} \right) . 
    \end{equation}
    The mean mass of a binary system in our simulated population is;
    \begin{equation}
    \bar{m}_{*, \mathrm{model}} = \frac{1}{\int_{m_{A}}^{m_{B}}f(m)dm} \int_{m_A}^{m_B} \int_{0}^1 f(m) g(q) (m+qm) \, dq dm 
        = \frac{3}{2}   \frac{1-\alpha_3}{2-\alpha_3} \left(\frac{m_{B}^{2-\alpha_3}-m_{A}^{2-\alpha_3}}{m_{B}^{1-\alpha_3}-m_{A}^{1-\alpha_3}} \right).
    \end{equation} 
    Thus, the total modeled mass $M_{*,\mathrm{model}}$ represents only a fraction of the total stellar population mass $M_*$:    
    \begin{equation}
    f_\mathrm{corr} = \frac{M_{*,\mathrm{model}}}{M_*} = f_{\mathrm{model}} \frac{\bar{m}_{*,\mathrm{model}}}{\bar{m}_{*}}
    \end{equation}
    and we must renormalize by the inverse of $f_\mathrm{corr}$ in order to return to the population we intended to simulate.

    Adopting the \citet[][]{10.1046/j.1365-8711.2001.04022.x} IMF, namely $\alpha_1=0.3$, $\alpha_2=1.3$, $\alpha_3=2.3$, $m_1=0.08\, \mathrm{M}_\odot$, $m_2=0.5\, M_\odot$, using the observed $f_\mathrm{bin}=0.7$ \citep[][]{2012Sci...337..444S}, arbitrarily choosing $m_\mathrm{min}=0.01\, \mathrm{M}_\odot$ and $m_\mathrm{max} = 200 \, \mathrm{M}_\odot$ as the minimum and the maximum stellar mass, and carrying out the simulation for primary masses in the range between $m_A=5\, \mathrm{M}_\odot$ and $ m_B= 150\, \mathrm{M}_\odot$ \citep[][]{2005Natur.434..192F}, we obtain $f_\mathrm{corr}^{-1}=4.78$.  
    

\section{Detection rate} \label{app:DetectionRate}

    To compute the detected BH merger rate by GW detectors, we follow a similar procedure to the one of \citet{2016ApJ...819..108B} which is a refined version of \citet{2015ApJ...806..263D}. In our cosmological calculation, we adopt the flat $\Lambda$CDM model with \mbox{$H_0=67.7 \, \frac{\text{km/s}}{\text{Mpc}}$} and $\Omega_m = 0.307$ \citep{2016A&A...594A..13P}. We follow the model of  \citet{2017ApJ...840...39M} for the star formation rate (SFR) model as a function of redshift, which is an updated version of \citet{2014ARA&A..52..415M},
    \begin{equation}
        \text{SFR}(z) = \frac{0.01 \cdot (1 + z)^{2.6}}{1 + ((1 + z) / 3.2)^{6.2}}  \, \, \text{M}_{\odot} \, \text{yr}^{-1} \, \text{Mpc}^{-3} \, .
    \end{equation}
    We assume that the metallicities of the binaries follow a truncated log-normal distributed,
    
    \begin{equation}
    \mathcal{N}(\log_{10}(Z) \, | \, \mu(z), \sigma) \equiv \frac{dP}{d \log_{10}(Z)}(z) = \frac{1}{\sigma \sqrt{2\pi}} \exp \left(- \frac{\left( \log_{10}(Z)-\mu(z) \right)^2}{2\sigma^2} \right)     
    \end{equation}
    with standard deviation $\sigma = 0.5$ and mean $\mu = \log_{10}\left(\bar{Z}(z)\right) - \frac{\ln(10)}{2} \sigma^2$ where the mean metallicity is \citep{2017ApJ...840...39M}
    \begin{equation}
        \bar{Z}(z) = Z_\odot \cdot 10^{0.153 - 0.074 \cdot z ^{1.34}} \, .
    \end{equation}
    The log-normal distribution is truncated at the highest metallicity bin edge, $Z_\mathrm{max} = 0.034923$, and the distribution is accordingly renormalized to ensure that $\int_{-\infty}^{\log_{10}Z_\mathrm{max}} \mathcal{N}(\log_{10}(Z) \, | \, \mu(z), \sigma)\, d\log_{10}Z = 1$. Portions of the distribution extending beyond the lower limit edge \mbox{$Z_\mathrm{min} = 0.000091$} are included in the edge bin when integrating over metallicity.
    
    We compute the detection rate by integrating the cosmological merger rate $R(z_m) = \frac{dN}{dm_1 dm_2 dV_c dt_m}$ per unit masses, per unit comoving volume, per unit time as measured in the source frame at the redshift of the merger as in Eq. (5) of \citet{2015ApJ...806..263D}, that is
    \begin{equation}
        R_{\rm det} = \iiint R(z_m) \frac{dt_m}{dt_{\rm det}} p_\mathrm{det} (z_m,m_{1},m_2) \, dm_1 dm_2 dV_c \, ,
        \label{eq:Rstart}
    \end{equation}
    where the factor $\frac{dt_m}{dt_{det}}=\frac{1}{1+z_m}$ account for the difference in clock rates at merger and at the detector and $p_\mathrm{det}$ is the detection probability accounting for the detector's selection effects. The integration over the comoving volume can be calculated with a change of variable over the redshift of merger, namely $dV_c = \frac{dV_c}{dz_m} d z_m$, where
    \begin{equation}
        \frac{dV_c}{dz_m}=\frac{4\pi c}{H_0} \frac{D^2_c(z_m)}{E(z)} \, \, ,
    \end{equation}
    and $D_c(z)$ is the comoving distance which is related to the luminosity distance as $D_L(z) = D_c(z) (1+z)$ and is computed as follows
    \begin{equation}
        D_c(z)=\frac{c}{H_0} \int_0^{z}\frac{dz'}{E(z')} 
    \end{equation}
    where $E(z) = \sqrt{\Omega_m (1+z)^3+\Omega_\Lambda}$.
    The merger rate $R(z_m)$ can be rewritten as the convolution of the star-formation rate, SFR$=\frac{d^2 M_f}{dV_c dt_f}$, that is the total mass of stars formed per comoving volume per year, and the number density of binaries per unit star forming mass $M_f$ per unit masses $m_1,m_2$ per unit log-metallicity interval $Z$ per unit time delay $\tau = t_m - t_f$:
    \begin{equation}
        R(z_m) \equiv R(z(t_m)) = \int^{t_m}_0 \int \frac{d^2 M_f}{dV_c dt_f} (z_f) \, 
        \frac{d^5 N}{dM_f dm_1 dm_2 d\tau d\log_{10}Z} (\tau) \, 
        \mathcal{N} \left( \log_{10}(Z) \, \bigg| \, \mu = \log_{10} \left( \bar{Z} (z_{f})\right) - \frac{\ln(10)}{2} \sigma^2  , \sigma = 0.5 \right)  \, d\log_{10}Z \, dt_f \, ,
    \end{equation}
    where we used the compact notation $z_f \equiv z(t_f)$. The time delay $\tau$ is mostly set by $T_\mathrm{merger}$, since the GW-driven merger takes much longer than stellar evolution for BH progenitors.  
    Performing the change of variable $dz_m = \frac{dz_m}{dt_m} dt_m = H_0 (1+z_m) E(z_m) dt_m$, the integral of Eq. (\ref{eq:Rstart}) translates into the following Monte-Carlo sum over the formation time intervals arbitrarily chosen as $\Delta t_i = 100 \, \text{Myr}$ and 30 
    uniformly-distributed log-metallicity intervals for $Z \in [Z_\mathrm{min},Z_\mathrm{max}]$
    \begin{equation}
        R_{\rm det} = \sum_{\Delta t_i} \sum_{\Delta Z_j} \sum_k \frac{\text{fSFR}(z_{f,i})}{M_{\mathrm{sim},\Delta Z_j}} \, f_\mathrm{corr} \, 4 \pi c\, D_c^2(z_{m,i,k})\, p_\mathrm{det}(z_{m,i,k}, m_{1,k}, m_{2,k}) \, \Delta t_i \, ,
    \end{equation}
    where  $M_{\mathrm{sim},\Delta Z_j}$ is the total mass simulated per log-metallicity interval $\Delta Z_j$ and fSFR is the total mass of stars formed per comoving volume per year per log-metallicity interval $\Delta Z_j$,
    \begin{equation}
    \begin{split}
        \text{fSFR}(z_{f,i}) &= \int_{\Delta Z_j} \text{SFR} \left(z_{f,i} \right) \, \mathcal{N} \left( \log_{10}(Z) \, \bigg| \, \mu = \log_{10} \left( \bar{Z} (z_{f})\right) - \frac{\ln(10)}{2} \sigma^2  , \sigma = 0.5 \right) d\log_{10}Z = \\
        & = \text{SFR}(z_{f,i}) \left[ CDF\bigg(\log_{10}(Z_j)+ \frac{\Delta Z_j}{2} \bigg) - CDF \bigg(\log_{10}(Z_j)-\frac{\Delta Z_j}{2} \bigg) \right] \, \, \text{M}_{\odot} \, \text{yr}^{-1} \, \text{Mpc}^{-3} \, ,
    \end{split}
    \end{equation}
    where $Z_j$ is the center of the log-metallicity bin $\Delta Z_j$ corresponding to the metallicity $Z_k$ of the binary $k$.  Meanwhile, the integrated SFR (iSFR) over the cosmic time used to obtain the weighted distributions of parameters after the CE phase is computed with the change of variable $dt = \frac{dt}{dz} dz = \left(H_0(1+z)E(z)\right)^{-1} dz$,
    \begin{equation}
        \text{iSFR}(Z) = \int_0^{\infty} \text{SFR}\left(z\right) \, \mathcal{N} \left( \log_{10}(Z) \, \bigg| \, \mu = \log_{10} \left( \bar{Z} (z_{f})\right) - \frac{\ln(10)}{2} \sigma^2  , \sigma = 0.5 \right) \frac{dt}{dz} dz   \, \, \, \, \text{M}_{\odot} \, \text{Mpc}^{-3} \, ,
    \label{eq:ISFR}
    \end{equation}
    which gives the total mass of stars formed per comoving volume at a given metallicity $Z$.

\section{Linear interpolation of the MESA simulations}\label{app:Interpolation}

    Running MESA simulations on the entire simulated binary population is too computationally  expensive. Instead, we use linear interpolation over a simulated grid to estimate the physical observables of the binaries that we are interested in. To generate the first simulations we sample stochastically in the logarithmic parameter space of initial masses, orbital periods and metallicities. We generate 3,000 initial points with $m_\mathrm{BH}\in [2.5\, \text{M}_\odot,60\, \text{M}_\odot]$, $m_\mathrm{He-star}\in[2.5\, \text{M}_\odot,89\, \text{M}_\odot]$, $p\in[0.05 \, \text{days},8.5 \, \text{days}]$ and \mbox{$Z\in[0.0001, 0.0349]$}.  We add to these 1,500 points drawn from a kernel density estimator (KDE) of the parameter distribution of the synthetic binary population. 
    
    We cover these points by running MESA binary simulations as described in section \ref{sec:MESA}.  We want to interpolate six quantities: the He-star mass and its CO core mass before the supernova, the resulting BH mass, the orbital period before the supernova, the lifetime of the BH-He-star binary and the spin of the second-born black hole. All physical quantities are log-transformed and rescaled to the interval $[-1,1]$ before going through the interpolation algorithm. The interpolation itself relies on building a Delaunay triangulation of the input data points followed by barycentric linear interpolation over the vortices of the (hyper)triangle containing the location of interest.  The relative error on each quantity $X_i$ is computed as $\Delta_i = (X_\mathrm{true,i} - X_\mathrm{interp.,i})/X_\mathrm{true,i}$ (in the original units, except for spins, where the relative error is computed on the spin logarithm to avoid excessive sensitivity to true values close to zero).  We then combine the relative errors of the quantities to obtain the combined relative error 
    \begin{equation}
        \Delta = \min \left( \sqrt{\sum_{i=1}^6 \Delta_i^2} \, , \sqrt{6} \right) \, ,
    \end{equation}
    where we arbitrarily limit the maximal combined relative error to $\sqrt{6}$, that is a point with all relative errors equal to $|\Delta_i| = 1$.

    We check the accuracy of the linear interpolator by conducting 50 leave 5\% of the sample out tests.  We use the combined relative errors as weights to sample an additional 500 points where the interpolator is performing the worst.  We iterate this procedure 21 times for a total of 10,500 simulations, stopping because almost all the new points generated through this procedure in the 22nd iteration would be on the boundaries.  The triangulation scheme can still fail near the parameter space boundaries; in this case, we find that 2.5\% of synthetic population systems cannot be interpolated, and we run 3000 simulations to bring the parameter space coverage to 100\%.
   
\begin{figure}
    \centering
    \includegraphics[width=\hsize]{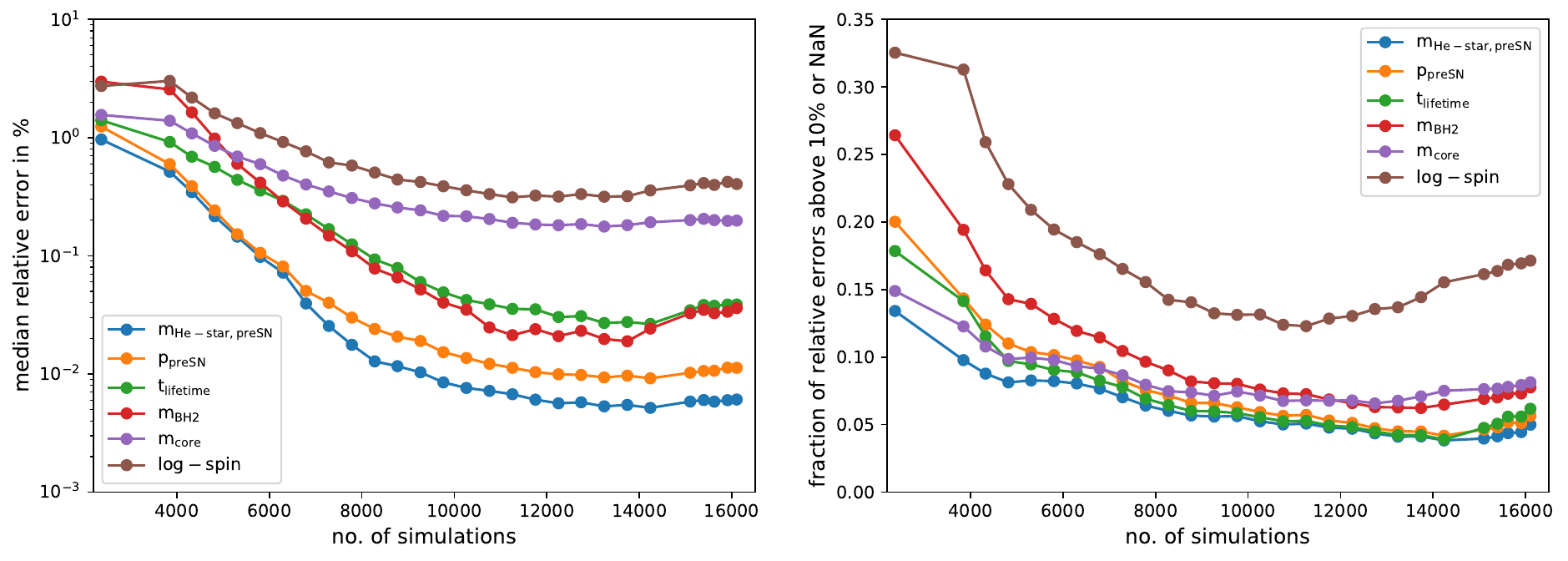}
        \caption{Median relative error of the interpolation expressed as a percentage (\textit{Left}) and the fraction of relative errors above 10\% (\textit{Right}) from all iterations of the 50 leave 5\% of the sample out tests for six interpolated quantities. The points on each plot, moving from left to right, represent the different iterations; we exclude all simulations that stopped due to initial Roche-lobe overflow (indicating a difference between COMPAS and MESA models). The right plot includes NaNs (obtained from non-fittable points, e.g., points at the boundary of the parameter space) when counting relative errors larger than 10\%, while the left plot excludes them.}
         \label{FigRelErr}
   \end{figure}

    Fig.~\ref{FigRelErr} shows the relative errors in the interpolated quantities over the series of 50 leave 5\% of the sample out tests.  The left panel shows the median percentage relative errors excluding non-fittable points.  These stabilize at 0.01\%,  0.20\%  and 0.04\% for the He-star mass before supernova, its CO core mass, and resultant BH mass; 0.01\% for the orbital period; 0.04\% for the lifetime of the BH-He-star binary and 0.4\% for the log-spin of the second-born BH. The log spin is the parameter which shows the biggest relative errors because it can have very large negative values for spins close to zero.  The right panel of Fig.~\ref{FigRelErr} shows the fraction of relative errors larger than 10\% as a function of the number of simulations for the different interpolated quantities where we also count non-fittable points, such as points at the boundary of the parameter space or isolated regions of the parameter space. At the last iteration the mean of the 50 leave 5\% of the sample out tests shows the following fraction of relative errors greater than 10\%: 5\% for the He-star mass before the supernova, 8\% for its CO core mass and the resulting BH mass, 6\% for the orbital period, 6\% for the lifetime of the BH-He-star binary and 17\% for the log-spin of the second-born BH.  The apparent increase in the fraction of relative errors larger than 10\% with the number of simulations happens because with the last iterations we are sampling mostly the boundaries of the parameter space and the test picks up more points that cannot be interpolated (we note that the median relative errors does not show this trend and stabilizes). The last simulations used to bring the coverage of the parameter space to 100\% are run in disconnected and remote regions of the parameter space, and the ``leave out'' tests pick up the newly added samples, artificially increasing the apparent fraction of large relative errors. 
    
\end{appendix}

\end{document}